\documentclass[aps,prx,reprint,noshowpacs,superscriptaddress,floatfix,letterpaper,longbibliography, notitlepage]{revtex4-2}
\usepackage{amsmath,amssymb,amsbsy,amsfonts,amsthm,bbm,bm,mathtools,mathrsfs}
\usepackage{color}
\usepackage{physics}
\usepackage{sidecap}
\usepackage{xfrac}
\usepackage[dvipsnames]{xcolor}
\definecolor{LapisLazuli}{RGB}{47, 102, 169}
\usepackage[colorlinks=true,citecolor=LapisLazuli,linkcolor=LapisLazuli,urlcolor=LapisLazuli]{hyperref}
\usepackage{empheq}
\usepackage{pgfplots}
\usepackage{stackengine}
\usepackage{relsize}
\usepackage[inline]{enumitem}
\usepackage[normalem]{ulem}
\usepackage{sidecap}
\usepackage{comment}
\usepackage{import}
\usepackage[english]{babel}
\usepackage[most]{tcolorbox}
\UseRawInputEncoding

\usepackage{array} 
\usepackage{booktabs} 
\usepackage{pgfplots}
\pgfplotsset{compat = newest}
\usetikzlibrary{arrows,intersections}
\usepackage{tikz-3dplot}
\usepackage{tikz}
\usepackage[mathscr]{euscript}
\graphicspath{{../plots/}}

\usepackage{pgfplots}
\pgfplotsset{compat = newest}
\usetikzlibrary{arrows,intersections}
\usepackage{tikz-3dplot}
\usepackage{tikz}
\newcommand{\HHquad}{\hspace{0.25em}} 

\makeatletter
\def\maketitle{
	\@author@finish
	\title@column\titleblock@produce
	\suppressfloats[t]}
\makeatother

\begin{document}

\title{Force Geometry and Irreversibility in Nonequilibrium Overdamped Dynamics}
\author{Erez Aghion}
\affiliation{Department of Physics, University of Louisiana at Lafayette, Lafayette,  LA 70504 , USA}
\email[]{erez.aghion@louisiana.edu}
\author{Swetamber~Das}

\affiliation{Department of Physics, School of Engineering and Sciences,\
	SRM University-AP, Amaravati, Mangalagiri 522240, Andhra Pradesh, India
}
\email[]{swetamber.p@srmap.edu.in}

\begin{abstract}
Recent experiments have revealed heterogeneous dissipation in optically trapped systems, often anticorrelated with local positional fluctuations, exposing a structural gap in the scalar stochastic thermodynamic description. While the scalar framework successfully quantifies dissipation through currents and entropy production rates, it does not reveal the underlying vectorial force geometry that shapes spatial dissipation patterns. Here, we bridge this gap by identifying force geometry as an organizing principle for nonequilibrium thermodynamics, introducing force alignment as a geometric determinant of irreversibility. We show that entropy production depends not only on force magnitudes but also on the relative orientation between deterministic driving forces and entropic gradients, vanishing only under exact anti-alignment with matched magnitudes. We formalize this geometric alignment through a time-dependent force-correlation coefficient, quantifying the relative orientation between the forces. This yields an instantaneous geometric lower bound on entropy production that remains valid even when force magnitudes are matched. For overdamped dynamics, perfect anti-alignment defines a thermodynamic stall where  net transport vanishes and the lower bound on entropy production is saturated. This force-level perspective provides a structural explanation for the experimentally observed fluctuation-dissipation anticorrelation and nonuniform dissipation. We construct geometric control charts for both constant dragging and sinusoidal driving protocols, explicitly locating experimental operating points within this force-space representation. Together, these results position force geometry as a unifying structural perspective on irreversibility, spanning active biological systems, microrheology, and naturally extending to underdamped dynamics. 
\end{abstract}

\maketitle

\section{Introduction}
Irreversibility is a defining feature of nonequilibrium dynamics, commonly quantified by finite entropy production. Its physical origin, however, is typically framed statistically rather than in terms of the forces that drive the dynamics~\cite{Sekimoto1998Langevin}. In stochastic thermodynamics, time-reversal asymmetry appears as probability currents, while the entropy production rate can also be expressed at the level of the underlying forces. Yet in practice, analysis proceeds primarily at the level of currents, treating dissipation as an aggregate quantity rather than resolving the organization of the contributing forces. In systems maintained out of equilibrium, entropy production arises from the interplay of competing forces, highlighting the need to understand how irreversibility is organized through force correlations, in space and time. 

Recent spatially resolved measurements provide a concrete setting in which the limitations of current-centric descriptions become evident. In particular, Ref.~\cite{DiTerlizzi2024} reported pronounced spatial inhomogeneity in entropy production across driven red blood cell membranes, together with counterintuitive anti-correlations between dissipation and positional fluctuations. Their observation revealed nonuniform heat dissipation along the cell contour, demonstrating that energetic cost can vary substantially across space even under steady driving. This spatial heterogeneity points to a missing structural layer in nonequilibrium thermodynamics, essential for understanding how dissipative processes are structured and regulated in living systems.

At the same time, these observations also expose a conceptual gap. While variance-based sum rules and force-free kinetic inference approaches enable experimental estimation of local entropy production~\cite{DiTerlizzi2024,DiTerlizzi2025}, they do not explain how entropy production is organized at the force level, or how the relative alignment of underlying forces shapes spatial patterns in dissipation. More broadly, established frameworks in stochastic thermodynamics -- including current-based formulations, thermodynamic uncertainty relations and Fisher-information speed limits -- provide powerful scalar bounds on entropy production, precision, and timescales~\cite{Barato2015Thermodynamic,Gingrich2016Dissipation,Seifert2019TUR,horowitz2020thermodynamic,Nicholson2020TimeInformation,DechantSasa2021PhysRevX11.041061}. However, these frameworks do not directly capture the geometric structure encoded in thermodynamic forces' vectorial alignment and its role in organizing dissipation. This necessary shift toward structural force alignment conceptually echoes recent bivectorial formulations of nonequilibrium thermodynamics, which establish that irreversibility fundamentally emerges from the cyclic coupling of flux and non-conservative forces in phase space~\cite{yang2021bivectorial,Byrne2023}.

Here, we show that irreversibility in overdamped diffusive systems admits a geometric interpretation at the force level, governed by the relative orientation between external driving forces and entropic (information) gradients. Entropy production vanishes only when these forces cancel exactly -- equal in magnitude and perfectly anti-aligned -- whereas any deviation in alignment or magnitude generates dissipation. We introduce an instantaneous force--correlation coefficient $r(t)$ that quantifies this alignment and provides an experimentally accessible measure of force organization. As analytically tractable example, we apply this framework to moving harmonic traps and diffusion on a ring in one and two dimensions,  constructing control charts that organize entropy production, transport, and driving strength across protocols. These charts provide a geometric account of the experimentally observed spatial heterogeneity and dissipation--fluctuation anti-correlation. Together, these results show that microscopic force geometry  sets  organizing principles in nonequilibrium thermodynamics,  complementing current-based bounds on irreversibility.

The paper is organized as follows. In Sec.~\ref{sec:force-geometry-entropy-production}, we define a net thermodynamic force as the sum of the external physical force and an information-theoretic force arising from the gradient of the surprisal, $-\ln \rho$. We then develop a force-based expression for the entropy production rate and identify a geometric contribution given by the inner product of these two forces. Based on this inner product, we introduce a force-correlation coefficient $r(t)$, which leads to a geometric hierarchy of irreversibility in overdamped systems and a finite-time bound on entropy production. Section~\ref{sec:moving-harmonic-trap} applies the force-geometry framework to a moving harmonic trap, considering both constant dragging and sinusoidal driving, and discusses connections to the RBC experiments of Ref.~\cite{DiTerlizzi2024}. In Sec.~\ref{sec:high-dimensional-angles}, we explore a two-dimensional harmonic trap and diffusion on a ring as examples extending the framework beyond one-dimensional Gaussian dynamics. In Sec.~\ref{sec:existing-thermodynamic-bounds}, we discuss the complementarity of this framework with existing thermodynamic bounds. The paper concludes in Sec.~VI with a discussion on force-geometric optimization at fixed mean transport and possible extensions to underdamped dynamics. Detailed derivations are given in the appendices.

\section{Force geometry and entropy production }
\label{sec:force-geometry-entropy-production}
\begin{figure}[t]
\centering
\includegraphics[width=0.6\columnwidth]{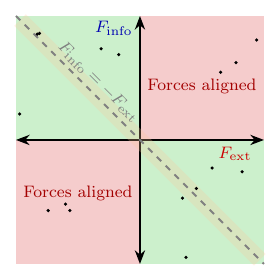}

\caption{ {\textbf{Force correlation diagram and scalar force geometry.}
Schematic representation of the alignment between the external force
\(F_{\mathrm{ext}}\) and the information-theoretic force
\(F_{\mathrm{info}}\) in one spatial dimension. The horizontal and vertical
axes denote the scalar force values \(F_{\mathrm{ext}}\) and
\(F_{\mathrm{info}}\), respectively, and each point represents a local force
pair \((F_{\mathrm{ext}}(x,t),F_{\mathrm{info}}(x,t))\) sampled from the system
at time \(t\). Green quadrants correspond to anti-aligned forces
\((F_{\mathrm{ext}}F_{\mathrm{info}}<0)\), while red quadrants correspond to
aligned forces \((F_{\mathrm{ext}}F_{\mathrm{info}}>0)\). The dashed diagonal
marks the exact cancellation condition \(F_{\mathrm{info}}=-F_{\mathrm{ext}}\),
for which the net thermodynamic force vanishes pointwise. The shaded band
schematically illustrates a concentration of force pairs near this cancellation
line, indicating strong but imperfect anti-alignment. In higher dimensions, the corresponding
force-correlation coefficient is defined using the \(\rho\)-weighted vector
inner product $\langle \vec F_{\mathrm{ext}} \cdot \vec F_{\mathrm{info}}\rangle$.}}
\label{fig:force_geometry}
\end{figure}

We consider overdamped stochastic dynamics, in $d\geq1$ dimensions, of  particles in contact with a heat bath at temperature $T$ and a potential landscape $U(\vec{x},t)$ (where $\vec{x}=(x_1,\dots x_d)$). These dynamics follow the set of Langevin equations:  
\begin{align}
    &\dot{x}_1=-\frac{1}{\gamma}\partial_{x_1}U(\vec{x},t)+\sqrt{2D}\Gamma_1(t), \nonumber\\ 
    &\dots \nonumber\\ 
    & \dot{x}_d=-\frac{1}{\gamma}\partial_{x_d}U(\vec{x},t)+\sqrt{2D}\Gamma_d(t). 
    \label{EqSetOfLangevins}
\end{align} 
Here, the diffusion coefficient $D>0$ is identical in every direction, $\gamma=k_BT/D$ and $k_B$ is Boltzmann's constant. The Gaussian white noise terms, represented by the $\Gamma_n(t)$s ($n=1,\dots d$), have zero mean and $\langle \Gamma_n(t)\Gamma_m(t')\rangle=
\delta_{mn}\delta(t-t')$.
The particles' spatial probability density, $\rho(\vec{x},t)$, expands through the Fokker--Planck equation~\cite{Risken1989Fokker}: 
\begin{align}
    {\partial_t}{\rho}(\vec{x}, t) = \frac{1}{\gamma} \vec{\nabla} \cdot \left( \rho(\vec{x}, t) \vec{\nabla} U(\vec{x}) \right) + D {\nabla}^2 {\rho}(\vec{x}, t). 
    \label{EqFP}
\end{align}

A central quantity characterizing nonequilibrium processes is the {internal} entropy production rate, which for this system reads~\cite{Seifert2005Entropy}: 
\begin{equation}
\dot{S}_i(t)
= \frac{k_B}{D} \left\langle \frac{|\vec{J}(\vec{x},t)|^2}{\rho^2(\vec{x},t)} \right\rangle, 
\label{EqEntropyProductionPrigongine}
\end{equation}
 where $\vec{J}(\vec{x},t)$ is the probability current, defined through {$\partial_t\rho(\vec{x},t)=-\vec{\nabla}\cdot\vec{J}(\vec{x},t)$}. Throughout this manuscript, angular brackets denote spatial averaging with respect to $\rho(\vec{x},t)$ and $|\cdot|$ represents vector magnitude.
Equation~\eqref{EqEntropyProductionPrigongine}  is consistent with the classical Prigogine formulation of entropy production in nonequilibrium thermodynamics~\cite{glansdorff1971thermodynamic}. 
In thermal systems, the heat dissipation rate (from the system to the bath) satisfies
$\dot{\mathcal Q}(t)=T[\dot S_i(t)-\dot S(t)]$,
where $T$ is the bath temperature and $\dot S(t)$ is the total rate of change
of system entropy. When steady-state heat balance is reached, the entropy production rate
and heat dissipation rate coincide \cite{Seifert2005Entropy}.

\begin{figure*}[t]
\centering
\begin{tcolorbox}[
    enhanced,
    colback=white,
    colframe=black,
    boxrule=0.8pt,
    arc=0pt,
    outer arc=0pt,
    leftrule=2pt,
    left=8pt,
    right=8pt,
    top=6pt,
    bottom=6pt,
    boxsep=0pt,
    width=0.92\textwidth,
    center title,
    before upper={\centering}
]
\textbf{Geometric hierarchy of irreversibility}

\vspace{0.8em}

{\color[RGB]{0,140,0} \textbf{Reversible condition: zero transport, zero entropy production}}\\
\textit{Local  force cancellation}: $\vec{F}_{\mathrm{info}}(x,t) = -\vec{F}_{\mathrm{ext}}(x,t)$\\
$r(t) = -1$, $\langle \vec{\mathcal{F}}_{\mathrm{net}} \rangle = 0$, $\dot{S}_i = 0$
\vspace{0.6em}

$\downarrow$

\vspace{0.2em}
\textit{Departing from this condition generically yields entropy production}

\vspace{0.6em}

{\color[RGB]{220,120,0} \textbf{Stall condition: zero mean transport, non-zero entropy production}}\\
\textit{Global (statistical) force anti-alignment}\\
$r(t) = -1$, $\langle \vec{\mathcal{F}}_{\mathrm{net}} \rangle = 0$, $\dot S_i > 0$\\
\vspace{0.6em}

$\downarrow$

\vspace{0.2em}
\textit{Departing from this condition produces dissipative finite transport}

\vspace{0.6em}

{\color[RGB]{180,0,0} \textbf{Driven nonequilibrium states}}\\
$r(t) > -1$,  $\langle \vec{\mathcal{F}}_{\mathrm{net}} \rangle \neq 0$, $\dot S_i > 0$

\vspace{1em}

\textbf{Geometric waste.} Entropy production can be decomposed into a mean-transport contribution and geometric waste arising from fluctuations of the residual net force. At thermodynamic stall, $r(t)=-1$, the mean transport contribution vanishes, while geometric waste $\propto {\rm Var}(\vec{\mathcal{F}}_{\rm net})$ can persist due to  local force imbalances. 

\end{tcolorbox}
\caption{\textbf{Conceptual hierarchy of force geometry in overdamped nonequilibrium dynamics.} Geometric waste is the total entropy production arising from departures from the
reversible force-cancellation condition. While equilibrium trivially satisfies  $r(t)=-1$, the converse is not true: global force anti-alignment does not preclude irreversible currents or finite entropy production. Also, see Appendix~\ref{Appen_force_coeff_stall} which discusses how $r = -1$ is related to thermodynamic stall.}
\label{box:geometric_hierarchy}
\end{figure*}

 {Using Eq.~(\ref{EqFP}), in Appendix~\ref{AppenDervEntrp} we derive an alternative representation of the entropy production,   by rewriting the entropy production rate in terms of the net thermodynamic force: 
\begin{align}
\dot{S}_i
&=
\frac{D}{k_B T^2}
\left\langle
|\vec{\mathcal{F}}_{\rm net}|^2
\right\rangle \nonumber\\
&=
\underbrace{
\frac{D}{k_B T^2}
\left|
\left\langle\vec{\mathcal{F}}_{\rm net}\right\rangle
\right|^2
}_{\text{transport contribution}}
+
\underbrace{
\frac{D}{k_B T^2}
{\rm Var}(\vec{\mathcal{F}}_{\rm net}),
}_{\text{geometric waste}}
\label{EqEPR}
\end{align} 
where
\begin{align}
 \vec{\mathcal{F}}_{\mathrm{net}} = \vec{F}_{\mathrm{ext}} + \vec{F}_{\mathrm{info}}.
 \label{EqNetForce}
 \end{align}
 {The first term on the RHS of Eq.~\ref{EqEPR} represents the transport contribution to $\dot S_i$ because it also appears in the mean transport velocity $\frac{d}{dt}\langle\vec{x}\rangle = \gamma^{-1}\langle \vec{\mathcal{F}}_{\mathrm{net}}\rangle$. The second term is the geometric contribution  due to variance ${\rm Var}(\vec{\mathcal{F}}_{\rm net})$ contains the $\rho$-wighted inner product $\langle \vec{F}_{\mathrm{ext}} \cdot \vec{F}_{\mathrm{info}}\rangle$.}}

This relation in Eq.~\eqref{EqNetForce} reveals a mechanistic perspective on
stochastic thermodynamics. Here, $\vec{F}_{\mathrm{ext}}$ denotes the externally
imposed force.  {For a conservative potential, $\vec{F}_{\mathrm{ext}}
=-\vec{\nabla} U$. More generally, however, the framework can be extended to systems where $\vec{F}_{\rm ext}$ is not strictly conservative and may have active force contributions, as long as the system is Markovian.}

We define the information-theoretic force as
\begin{equation}
\vec{F}_{\mathrm{info}}
=
-k_B T \vec{\nabla} \ln \rho .
\label{EqFInfo}
\end{equation}
This force is an entropic restoring force arising from spatial variations of
the probability density, and acts locally to oppose spatial probability
gradients. In the absence of nonequilibrium driving, it leads to relaxation
toward the equilibrium distribution by counteracting local probability
inhomogeneities. At equilibrium, the information-theoretic force $\vec{F}_{\mathrm{info}}$
exactly balances the mechanical force $\vec{F}_{\mathrm{ext}}$, whereas under
external driving the two forces compete to determine the structure of entropy
production.

While mathematically consistent with standard current-based expressions for the
entropy production rate~\cite{Seifert2005Entropy}  {(see also Appendix
Table~\ref{tab:velocity-force-representations})}, Eq.~\eqref{EqNetForce}
exposes a geometric structure that is otherwise obscured in scalar observables
and information-theoretic characterizations of nonequilibrium dynamics.

Through Eq.~\eqref{EqNetForce} we find a geometric constraint on reversibility arising from the force decomposition. 
The entropy production rate vanishes only when the two forces are perfectly anti-aligned and equal in magnitude, $\vec{F}_{\mathrm{info}}(\vec{x},t) = -\vec{F}_{\mathrm{ext}}(\vec{x},t)$, for any $\vec{x}$.
Anti-alignment is a necessary condition for reversibility; matching force magnitudes alone is not sufficient. 
In other words, equilibrium requires both directional and magnitude cancellation, whereas nonequilibrium arises from failure of either, independently of whether the system is in a steady state or undergoing transient dynamics. Within the present force-based representation, irreversibility corresponds to a geometric asymmetry between mechanical and entropic contributions in local thermodynamic force space.

Expanding the square explicitly yields:
\begin{align}
\dot{S}_i(t)
&= \frac{D}{k_B T^2}
\left\langle |\vec{F}_{\mathrm{info}} + \vec{F}_{\mathrm{ext}}|^2 \right\rangle_t \nonumber \\ 
&= \frac{D}{k_B T^2}
\left[
\langle |\vec{F}_{\mathrm{info}}|^2 \rangle_t
+ \langle |\vec{F}_{\mathrm{ext}}|^2 \rangle_t
+ 2 \langle \vec{F}_{\mathrm{info}}\cdot\vec{F}_{\mathrm{ext}} \rangle_t
\right].
\label{eq:force_decomposition}
\end{align}
The entropy production rate therefore consists of three contributions:
(i) an information cost associated with maintaining nonequilibrium probability gradients,
(ii) a driving cost due to external forcing,
and (iii) a correlation term capturing constructive or destructive interference between forces.
Because this structure follows directly from the generic form of thermodynamic
forces in overdamped Langevin dynamics, it therefore does not depend on confinement
or linear response assumptions.
The information-theoretic force variance $\langle |\vec{F}_{\mathrm{info}}|^2\rangle$ is directly proportional to the Fisher information of the instantaneous probability distribution, linking this contribution to statistical measures of distinguishability.
By contrast, the force correlation term, often not emphasized  in scalar descriptions,
encodes the genuinely geometric organization of entropy production at the force level.
A schematic description of the forces in $(\vec{F}_{\mathrm{ext}},\vec{F}_{\mathrm{info}})$ force space is shown in Fig.~\ref{fig:force_geometry} in one spatial dimension, as illustration.

 {For fixed force variances, $\langle |F_{\mathrm{ext}}|^2\rangle$ and
$\langle |\vec{F}_{\mathrm{info}}|^2\rangle$, and fixed transport and noise parameters
($D$, $\gamma$, $T$), the entropy production rate is \textit{not} uniquely determined.
Instead, it depends explicitly on the correlation term
$\langle \vec{F}_{\mathrm{ext}}\cdot\vec{F}_{\mathrm{info}}\rangle$, which measures the relative spatial 
orientation between the two forces.
As a result, two systems with identical current-based observables and identical
fluctuation scales can dissipate very differently, depending solely on their force
geometry. This degree of freedom is invisible in scalar descriptions based on currents,
variances, or state-space distances, but plays a central role in shaping
irreversibility.}

\begin{figure*}[t]
\centering
\begin{minipage}[t]{0.52\textwidth}
    \centering
    \includegraphics[width=\linewidth]{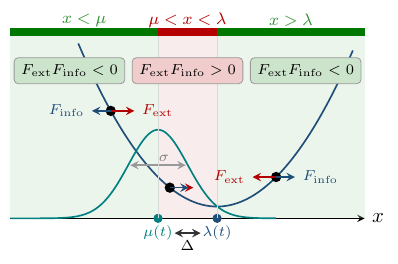}
\end{minipage}
\begin{minipage}[t]{0.47\textwidth}
    \vspace{-14.0em}
    \centering
    \normalsize
    \setlength{\tabcolsep}{8pt}
    \renewcommand{\arraystretch}{1.6}
    \begin{tabular}{@{}c c c c@{}}
    \hline
    Region & Force alignment & $\theta(x,t)$ & $\cos\theta(x,t)$ \\
    \hline
    $x<\mu$ & Anti-aligned & $\pi$ & $-1$ \\
    $\mu<x<\lambda$ & Aligned & $0$ & $+1$ \\
    $x>\lambda$ & Anti-aligned & $\pi$ & $-1$ \\
    \hline
    \end{tabular}
\end{minipage}
\caption{
\textbf{Force correlation regimes in a moving harmonic trap.}
Left: schematic illustrating force configurations in a  harmonically trapped particle, in  one dimension, 
shown in terms of the particle position $x$, trap center $\lambda$, and mean
position $\mu$.
Right: local angular dependence of the force alignment.
Anti-aligned and aligned configurations of the external and information-theoretic
forces correspond to negative and positive force correlations, respectively.
The ratio $|\Delta|/\sigma$ characterizes the force-correlation regime.}
\label{fig:harmonic_trap_schematic}
\end{figure*}

The natural dimensionless quantity emerging from
Eq.~\eqref{eq:force_decomposition} is the instantaneous non-centered Pearson correlation
coefficient between the external and information forces:
\begin{equation}
r(t)
= \frac{\langle \vec{F}_{\mathrm{info}}\cdot\vec{F}_{\mathrm{ext}} \rangle_t}
{\sqrt{\langle |\vec{F}_{\mathrm{ext}}|^2 \rangle_t \,
        \langle |\vec{F}_{\mathrm{info}}|^2 \rangle_t}}. 
        \label{EqPearsonInDDimensions}
\end{equation}
 {Here, the force correlation term is a function of the spatial angle, $\theta(\vec{x},t)$, between the forces:
\begin{align}
       &\langle \vec{F}_{\mathrm{info}}\cdot\vec{F}_{\mathrm{ext}} \rangle_t=\left\langle |\vec{F}_{\mathrm{info}}|\HHquad|\vec{F}_{\mathrm{ext}}|\cos[\theta(\vec{x},t)]\right\rangle_t
\end{align}}
The correlation coefficient, $r(t)$, quantifies the geometric alignment of fluctuating forces and
serves as a real-time indicator of irreversibility at the force level. Importantly, both $\dot S_i(t)$ and $r(t)$ are defined instantaneously, without assuming that the system is at steady-state or cycle-averaged, so force geometry and stall apply equally to steady, periodic, and
transient nonequilibrium dynamics.

 {{{Local orthogonality, $\vec F_{\rm ext}\cdot \vec F_{\rm info}=0$, identifies points where the geometric contribution to entropy production vanishes, so the local entropy produciton is determined solely by the magnitudes of the two forces. If they happen to be perpendicular in entire space, then the coefficient $r$ vanishes; otherwise, it is bounded $-1\leq r(t) \leq 1$.} Perfect force anti-alignment, when the relative angle between the external and information-theoretic forces is $\pi$ everywhere at every point in space, yields a force correlation coefficient $r(t) = -1$. This situation}  defines the
\textit{stall condition} of overdamped dynamics, where correlated cancellation of external and
information-theoretic forces leads to a vanishing \textit{mean} net force,
$\langle \vec{\mathcal{F}}_{\mathrm{net}} \rangle = 0$, and arrests macroscopic transport.
Yet such thermodynamic stall---a nonequilibrium steady state---does not imply reversibility.
Local force imbalances, $\vec{F}_{\mathrm{ext}}(\vec{x},t) \neq
-\vec{F}_{\mathrm{info}}(\vec{x},t)$, generally persist, for example in a
one-dimensional periodic steady state, sustaining local probability currents
and finite entropy production. In fact, at stall,
$\dot S_i = (D/k_BT^2)\mathrm{Var}(\vec{\mathcal{F}}_{\mathrm{net}})$, showing that entropy production
is entirely sustained by fluctuations of the net force $\vec{\mathcal{F}}_{\mathrm{net}}$. 
While residual local force imbalance may not be directly accessible in coarse-grained
experiments, its presence is thermodynamically unavoidable whenever stall does not coincide
with reversibility. The box in Fig.~\ref{box:geometric_hierarchy} summarizes this conceptual hierarchy.}

True thermodynamic reversibility is recovered only under the stricter, pointwise condition $\vec{F}_{\mathrm{info}}(\vec{x}) = -\vec{F}_{\mathrm{ext}}(\vec{x})$, which eliminates all probability currents and renders microscopic trajectories statistically time-symmetric. Any departure from this perfect local force cancellation---whether through force misalignment or unequal magnitudes---necessarily generates entropy production. 
 {We define \textit{geometric waste} as the contribution to entropy production arising from fluctuations of the residual net force about its mean, $\dot{S}_i^{\rm gw}=(D/k_BT^2)\mathrm{Var}(\vec{\mathcal{F}}_{\mathrm{net}})$.} It quantifies dissipation associated with spatially heterogeneous and incomplete force cancellation that does not contribute to mean transport. At thermodynamic stall, $r(t)=-1$ and the mean transport vanishes, so the entire entropy production is geometric waste. In this sense, $r(t)$ quantifies proximity to stall, whereas geometric waste distinguishes thermodynamic stall from true reversibility: a stalled state may have $\dot{S}_i^{\rm gw}>0$ because of residual local force imbalances, while true reversibility requires pointwise force cancellation and hence $\dot{S}_i^{\rm gw}=0$. Because $r(t)$ captures only global force alignment, these residual local imbalances are not fully resolved by $r(t)$ only.

 {Crucially, this framework highlights a geometric route to reducing entropy
production that is distinct from the quasistatic limit. In quasistatic or
near-reversible operation, entropy production is reduced by suppressing the
dynamics. Here, instead, entropy production is organized by the relative
alignment of the external and information-theoretic forces. At a given
mean net  force $\langle\vec{\mathcal{F}}_{\rm net}\rangle$, enhancing force anti-alignment can reduce the fluctuating
component of the net force and thereby lowers the corresponding
entropy production. The coefficient $r(t)$ provides a global geometric measure
of this alignment, relating the entropy production rate to the covariance
between the external and information-theoretic forces. In overdamped dynamics,
the condition $r(t)=-1$ identifies a stall condition, distinct from the
stronger condition of true reversibility, which requires pointwise cancellation
of the net thermodynamic force.}

\subsection{A tunable geometric lower bound on entropy production}

The force decomposition yields a natural geometric constraint on entropy production.
Using the inequality
$\langle |\vec{F}_{\mathrm{ext}}+\vec{F}_{\mathrm{info}}|^2\rangle_t
\ge (\sqrt{\langle |\vec{F}_{\mathrm{ext}}|^2\rangle_t}
-\sqrt{\langle |\vec{F}_{\mathrm{info}}|^2\rangle_t})^2$,
we obtain the instantaneous bound:
\begin{equation}
\dot S_i(t) \ge
\frac{D}{k_B T^2}
\left(
\sqrt{\langle |\vec{F}_{\mathrm{ext}}|^2\rangle_t}
-
\sqrt{\langle |\vec{F}_{\mathrm{info}}|^2\rangle_t}
\right)^2 . 
\label{EqInstanteneousInequality}
\end{equation}
The bound is saturated when the two forces are perfectly anti-aligned
($r(t) = -1$), corresponding to the stall condition of zero mean net force.
As shown in Fig.~\ref{box:geometric_hierarchy},  except for the special case of exact pointwise force cancellation, this lower
bound remains greater than zero.
Reversible stochastic operations ($\dot S_i = 0$) require the additional condition
that the force magnitudes are matched locally,
$\vec{F}_{\mathrm{info}}(\vec{x},t) = -\vec{F}_{\mathrm{ext}}(\vec{x},t)$.
Notably, unlike many thermodynamic bounds whose saturation requires fine-tuned or
asymptotic conditions and lacks a clear operational interpretation, saturation of the
present geometric bound corresponds directly to a physically meaningful regime—namely,
the stall condition of zero mean transport.
An integrated version of this bound is derived in  Appendix~\ref{AppenIntegratedBounds}.

Below, we use examples in one and two dimensions, to demonstrate the significance of force geometry. We begin with one dimension, which connects directly to two experimental set-ups, then show the example of two dimensional diffusion on a ring.

\section{The example of A moving harmonic trap in one dimension, and application to experiments}
\label{sec:moving-harmonic-trap}

\begin{figure*}[t]
\centering
 \includegraphics[width=.85\textwidth]{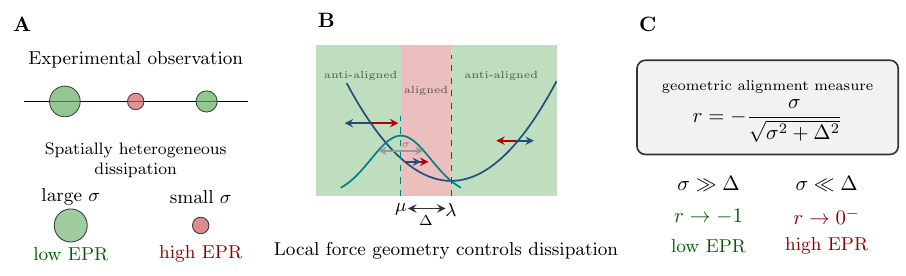} 
\caption{
\textbf{Geometric interpretation of experimental observations.}
Schematic illustrating how spatially heterogeneous  entropy production and
fluctuation--dissipation trends reported by~\cite{DiTerlizzi2024}
{may arise from} local force geometry.
(A) Spatial variation of inferred entropy production across coarse-grained membrane patches
(circle size schematically indicates positional variance).
(B) Local force configurations in a harmonic trap, highlighting spatial variation in force geometry.
(C) Force--correlation coefficient $r(t)$ as a geometric alignment measure linking positional variance
and entropy production.
}

\label{fig:schematic_exp_theory}
\end{figure*}
The moving harmonic trap provides a uniquely powerful setting in which we  translate force geometry
into experimentally testable design principles.
From a theoretical perspective, harmonic confinement is one of the few nonequilibrium stochastic
systems that is exactly solvable far from equilibrium, allowing explicit relations between forces,
fluctuations, and entropy production to be derived without severe approximations
\cite{Jarzynski1997Nonequilibrium,Seifert2012Stochastic}.
Experimentally, the same model underlies a broad class of biophysical and microrheological
measurements: optical and magnetic tweezers impose effective harmonic potentials on tracer
particles, membrane patches, and biomolecular assemblies, while controlled protocols such as
constant-velocity dragging or sinusoidal modulation are used to probe entropy production and mechanical
response with high spatiotemporal resolution
\cite{Neuman2004OpticalTrapping,Kumar2021Multi}.
In this setting, experimentally accessible observables including positional lag and variance map
 onto thermodynamic forces, making the harmonic trap an ideal platform for illustrating
the framework, identifying operating regimes, and constructing geometric control charts for
nonequilibrium systems.

In one dimension, we remove all vector notations, for simplicity. The harmonically trapped particle, described by the potential
$U(x,t) = \tfrac{1}{2}k\bigl(x-\lambda(t)\bigr)^2$, where $k$ is the trap stiffness and
$\lambda(t)$ the time-dependent trap centre, provides a canonical setting in which the
geometric structure of entropy production can be analyzed exactly.
Directed transport arises from the lag
$\Delta(t) = \mu(t) - \lambda(t)$ between the mean particle position $\mu(t)$ and the
trap centre.
Together with the positional variance $\sigma^2(t)$, both quantities are readily accessible
in experiments, making the harmonic trap an ideal bridge between force geometry and
measurable entropy production.
The detailed derivation of the force decomposition and correlation coefficient is provided
in Appendix~\ref{AppenGeneralFrameworkForMovingHarmonicTraps}. 

\subsection{Exact force decomposition and correlation}

The geometric origin of entropy production and its spatial heterogeneity becomes explicit when we examine the instantaneous forces acting on a particle in a harmonically trapped system driven by time-dependent protocols. For definiteness, we take $\Delta<0$ in Fig.~\ref{fig:harmonic_trap_schematic}\,, but this sign choice is purely conventional and leaves the classification of force-correlation regions unchanged. Overdamped dynamics in a harmonic potential is a linear Ornstein–Uhlenbeck process with additive Gaussian noise. Hence, starting from an initially Gaussian shape around $x=0$, the particles' spatial distribution remains Gaussian at all times, with time-dependent mean and variance~\cite{Risken1989Fokker}. Accordingly, the external driving force and the information-theoretic restoring force are both linear in position, but are centered on distinct reference points (see Fig.~\ref{fig:harmonic_trap_schematic}):
\begin{align}
F_{\mathrm{ext}}(x,t) &= -k\bigl(x-\lambda(t)\bigr), \qquad \nonumber\\
F_{\mathrm{info}}(x,t) &= \frac{k_B T}{\sigma^2(t)}\bigl(x-\mu(t)\bigr),
\end{align}
where $\lambda(t)$ denotes the trap center and $\mu(t)$ the mean particle position. {Here $\sigma(t)$ is the positional standard deviation, with $\sigma^2(t)=\langle[x-\mu(t)]^2\rangle_t$.} At each instant, $F_{\mathrm{info}}$ is directed outward from the mean 
$\mu(t)$, while $F_{\mathrm{ext}}$ is directed inward toward the trap center $\lambda(t)$. A finite lag
$\Delta(t) := \mu(t)-\lambda(t)\neq 0$ signals nonequilibrium transport.\\

Crucially, the local scalar product of the two forces:
\begin{equation}
F_{\mathrm{ext}}F_{\mathrm{info}}
= -\frac{k k_B T}{\sigma^2(t)}\bigl(x-\lambda(t)\bigr)\bigl(x-\mu(t)\bigr),
\end{equation}
changes sign across space: it is negative when $x$ lies outside the interval $[\lambda,\mu]$ and positive within it. This sign reversal reflects local constructive or destructive interference between driving and entropic restoring forces, thereby generating spatial heterogeneity in entropy production.\\

The force statistics follow exactly,
\begin{align}
\langle F_{\mathrm{ext}}^2 \rangle &= k^2(\Delta^2(t)+\sigma^2(t)), \nonumber\\
\langle F_{\mathrm{info}}^2 \rangle &= \frac{(k_B T)^2}{\sigma^2(t)}, \nonumber\\
\langle F_{\mathrm{ext}}F_{\mathrm{info}} \rangle &= -k k_B T,
\end{align}
leading to a compact expression for the non-centered Pearson correlation coefficient, Eq.~\eqref{EqPearsonInDDimensions}, which now assumes the form:
\begin{equation}
r(t)= -\frac{\sigma(t)}{\sqrt{\Delta^2(t)+\sigma^2(t)}}. 
\label{EqPearsonCorrelationCoefficient}
\end{equation}

This result shows that for diffusion in driven one dimensional harmonic potentials, force anti-alignment is controlled neither by the lag $\Delta$
nor by the fluctuations $\sigma$ alone, but by their dimensionless ratio
$|\Delta|/\sigma$.
When fluctuations dominate ($|\Delta|\ll\sigma$), the forces approach perfect
anti-alignment ($r(t)\to-1$) and entropy production is reduced. 
Conversely, when lag dominates ($|\Delta|\gg\sigma$), force alignment weakens
($|r(t)|\ll1$) and irreversibility increases.
The spatial organization of driven transport is thus captured by a single geometric
parameter that can be evaluated instantaneously from the force statistics.

\subsection{{A possible geometric interpretation of RBC observations}}
 {A useful experimental testbed for assessing the force-geometry framework developed here is provided by the colloidal and membrane-fluctuation experiments reported in Ref.~\cite{DiTerlizzi2024}. That work considers several optically controlled protocols, including a moving harmonic trap, a  stochastic switching harmonic trap, and a static-trap active-particle model for red blood cell (RBC) membrane fluctuations. Two observations are especially relevant for the present discussion. First, regions with large positional fluctuations can dissipate weakly, whereas comparatively quiescent regions can act as entropy-production hotspots, indicating an anticorrelation between positional fluctuations and entropy production. Second, the entropy production is spatially heterogeneous: even in steady operation, some regions dissipate strongly while others dissipate weakly.}

 {
Motivated by these observations, 
our aim is to ask whether the force geometry provides a mechanism consistent with the reported fluctuation--dissipation anticorrelation and spatially heterogeneous entropy production. In this interpretation, regions or regimes in which the external and information-theoretic forces are closer to anti-alignment can exhibit reduced entropy production, while departures from this force balance generate localized dissipation. 
}

In a driven harmonic trap, the dimensionless lag-to-width ratio $|\Delta|/\sigma$ provides a  geometric control parameter for the force--force correlation coefficient $r$.
Regions with poor force alignment ($|r|\ll1$, corresponding to $|\Delta|\gg\sigma$)
necessarily generate strong entropy production, whereas regions approaching force
anti-alignment ($r\approx-1$, or $|\Delta|\ll\sigma$) dissipate minimally.
As a result, positional fluctuations alone are not generally predictive of  entropy production: patches
with small fluctuations can be highly dissipative if their response lags the drive,
while strongly fluctuating regions can operate near force cancellation and dissipate
weakly.

This force-level perspective provides a  geometric framework of the experimentally
observed anti-correlation between variance and entropy production, as well as the
spatial heterogeneity characterized in the same measurements.
Rather than reflecting anomalous fluctuations or proximity to equilibrium, the apparent efficiency of certain membrane regions might emerge from spatially varying force anti-alignment (see Fig.~\ref{fig:harmonic_trap_schematic}), which would buffer the energetic cost of active fluctuations.
From this perspective, spatial heterogeneity in entropy production may not reflect
heterogeneous activity or noise intensity, but spatial variation in force geometry. A visual summary of this framework is shown in
Fig.~\ref{fig:schematic_exp_theory}.
Force geometry thus complements  force-free inference schemes~\cite{DiTerlizzi2025}, providing a
fundamental organizational framework that can guide the interpretation and design of
experiments probing nonequilibrium  entropy production.

The above considerations apply generally to harmonic traps driven by
time-dependent protocols $\lambda(t)$.
To illustrate how the force-geometry framework organizes nonequilibrium operation,
we consider two experimentally relevant driving protocols: constant-velocity
dragging and sinusoidal driving.
For each case, explicit relations between the force correlation $r$ and the driving
parameters can be obtained, enabling the construction of geometric representations
that compare  entropy production, transport, and force alignment across protocols.

\begin{figure*}[t]
\centering
\includegraphics[width=0.75\textwidth]{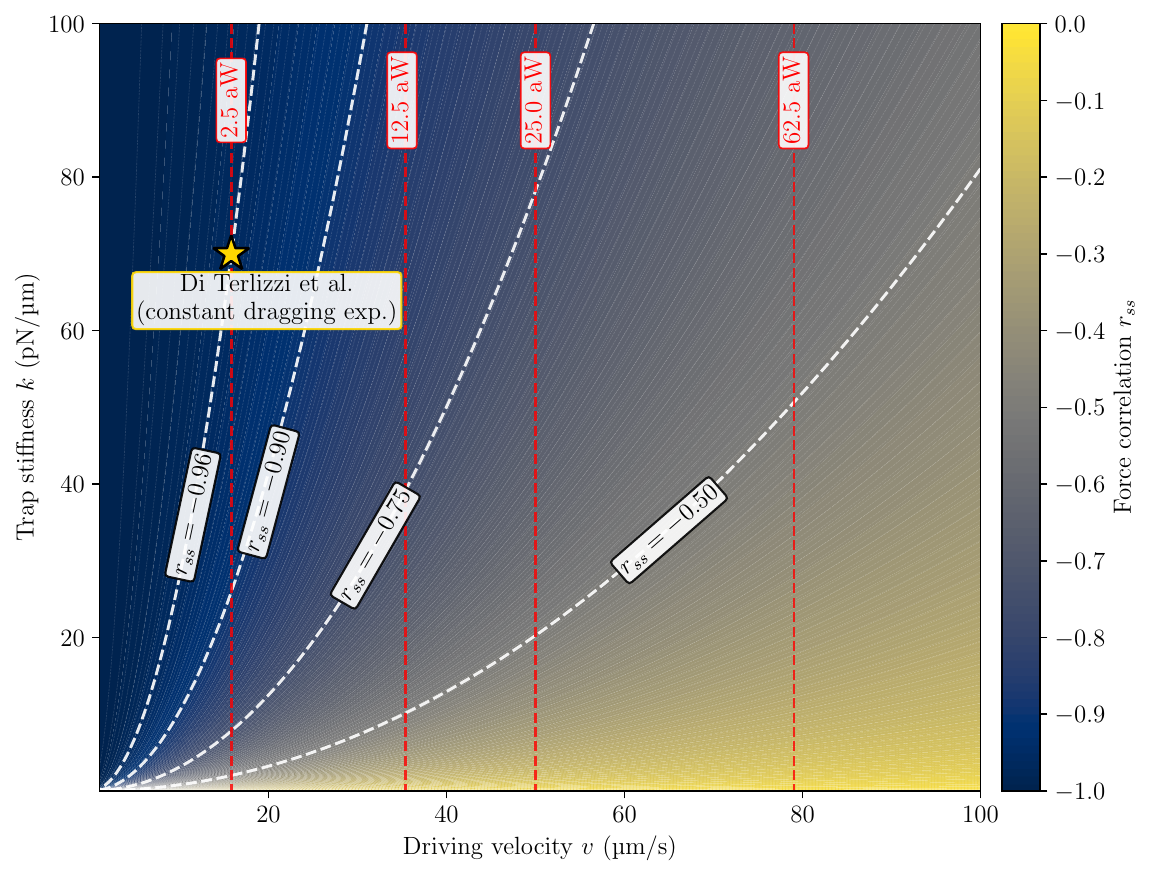}
\caption{
\textbf{Geometric organization under constant dragging.}
 {Heat map of the steady-state force--correlation coefficient $r_{\mathrm{ss}}$ in the
$(k,v)$ parameter space of a harmonically trapped particle, computed at fixed friction
$\gamma = 0.01\,\mathrm{pN\cdot s}/\mu\mathrm{m}$.
Dashed white curves denote contours of constant force correlation, while dashed red curves
indicate contours of constant injected power $P=\gamma v^2$.
Several representative correlation contours highlight regimes of strong force
anti-alignment at finite power.
The rescaled experimental operating point of Ref.~\cite{DiTerlizzi2024} {(for their constant dragging experiment)},
marked by a yellow star, lies within this strong anti-alignment region at the intersection of
$r_{\mathrm{ss}}\approx -0.96$ and $P \approx 2.5\,\mathrm{aW}$.
The lack of parallelism between iso-power contours and iso-correlation contours shows that
protocols with the same energetic cost can realize different force geometries.}
}
\label{fig:design_chart}
\end{figure*} 
\subsection{Geometric organization and experimental operating regimes}

For constant-velocity driving, $\lambda(t)=vt$, the system reaches a nonequilibrium
steady state characterized by
$\sigma = \sqrt{k_B T/k}$ and $\Delta = -\gamma v/k$ (see Appendix~\ref{AppenGeneralFrameworkForMovingHarmonicTraps} for derivation and Appendix~\ref{AppenExpMapping} for mapping to the experimental data).
The corresponding steady-state force correlation coefficient $r_{\rm ss}$, where
\begin{align}
r_{\mathrm{ss}} &= -\left[1+\left(\frac{\gamma v}{\sqrt{k k_B T}}\right)^2\right]^{-1/2}.
\label{eq:r_exact_const}
\end{align} 

This expression defines a simple geometric relation between force alignment and the
driving parameters.
Operating points with $r_{\mathrm{ss}}$ closer to $-1$ correspond to stronger force
anti-alignment.
Because $\sigma$ and $\Delta$ can be tuned independently through the trap stiffness
and driving velocity, the steady-state correlation provides a compact representation
for comparing nonequilibrium operating regimes in experiment.

 {We can write the steady state entropy production rate $\dot S^{\rm ss}_i$ for this model using $r_{\rm ss}$ as follows:
\begin{align}
\dot S^{\rm ss}_i = \frac{\gamma v^2}{T} = \frac{k k_B}{\gamma}
\left(
\frac{1}{r_{\rm ss}^2}-1
\right).
\end{align}
Thus, within this model, the low-dissipation limit corresponds to the strongly anti-aligned limit $r_{\rm ss}\to -1$ which is fluctuation-dominated $|\Delta_{\rm ss}|/\sigma_{\rm ss} \ll 1$, while the drive-dominated regime ($|\Delta_{\rm ss}|/\sigma_{\rm ss} \gg 1$) with $r_{\rm ss}\to0^-$ is highly dissipative. Thus, reducing the driving speed $v$ moves the system toward the anti-aligned configuration or the fluctuation-dominated regime and simultaneously lowers the entropy production rate, $\dot S_i\propto v^2$. }

Figure~\ref{fig:design_chart} presents a geometric control chart in the experimentally accessible parameter plane spanned by the trap stiffness $k$ and the dragging velocity $v$. The color scale encodes the steady-state force-correlation coefficient $r_{\mathrm{ss}}$, while dashed contours indicate constant values of $r_{\mathrm{ss}}$. The vertical dashed lines denote constant injected power, $P=\gamma v^2$,  which is fixed by the dragging velocity in the steady constant-drag protocol. Thus, at fixed steady-state transport speed $v$, changing the trap stiffness does not change the injected power or the steady-state entropy production rate, $\dot S_i^{\rm ss}=P/T$. Instead, $k$ acts as a geometric control parameter that changes how the same net thermodynamic force is decomposed into external and informational forces. For the steady constant-drag harmonic trap, the chart separates the dependence of energetic cost, controlled by $v$, from the dependence of force geometry, controlled by the dimensionless ratio $\gamma^2v^2/(k k_BT)$.

At steady state, Eq.~\eqref{eq:r_exact_const} shows that contours of fixed $r_{\mathrm{ss}}$ correspond to keeping $\frac{\gamma^2 v^2}{k k_BT}$
constant. Hence fixed-correlation contours obey the scaling $k\propto v^2$, which explains the parabolic structure of the contour lines in Fig.~\ref{fig:design_chart}. Maintaining a fixed degree of force anti-alignment while increasing the dragging speed therefore requires the trap stiffness to increase quadratically with $v$. The chart should therefore be interpreted as a map of protocol-dependent force geometry rather than as a claim that stiffness tuning lowers dissipation at fixed transport. It shows how protocols with the same injected power can realize distinct force decompositions, and how moving across power contours changes both the transport speed and the thermodynamic cost.

\begin{figure*}[t]
\centering
\includegraphics[width=0.75\textwidth]{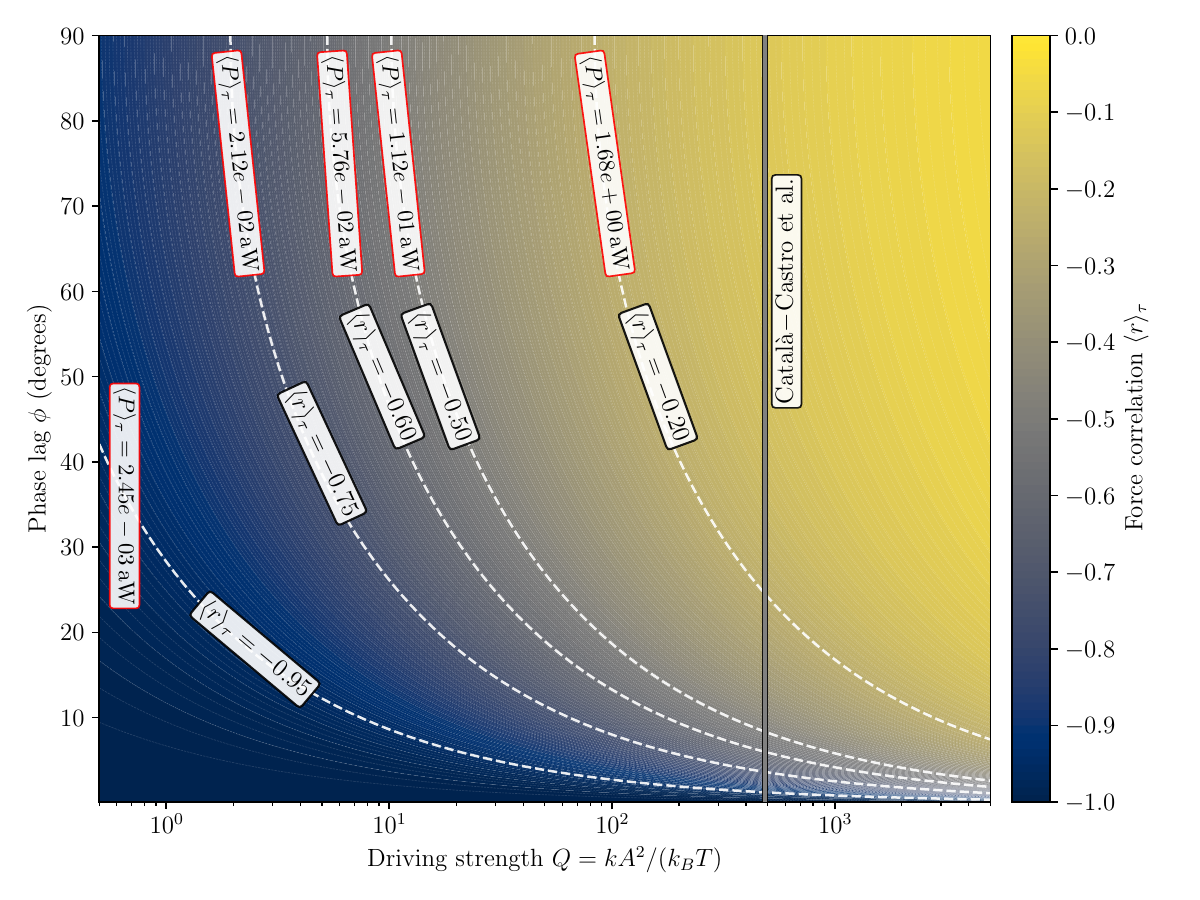}
\caption{
\textbf{Geometric organization under sinusoidal driving.}
 {Heat map of the time-averaged force--correlation coefficient $\langle r\rangle_\tau$
shown as a function of the dimensionless driving strength
$Q=kA^2/(k_BT)$ and phase lag $\phi=\tan^{-1}(\gamma\omega/k)$.
Dashed curves denote contours of constant force correlation, which coincide with
contours of constant time-averaged injected power $\langle P  \rangle_\tau$ for sinusoidal driving,
reflecting the shared underlying force geometry.
The vertical line indicates the driving strength $Q_{\mathrm{ext}}=486.6$ corresponding to 
the operating point used in the optical-tweezer microrheology experiments of Ref.~\cite{CatalaCastro2025} 
which lies predominantly in the high entropy production regime. 
}}
\label{fig:sinusoidal_design_chart}
\end{figure*}

To connect the geometric chart to experimental parameters, we mark the operating point corresponding to the constant-velocity dragging measurements reported in Ref.~\cite{DiTerlizzi2024}. Since that experiment used a different friction coefficient, $\gamma_{\mathrm{exp}}=0.025\,\mathrm{pN\,s}/\mu\mathrm{m}$, we rescale the driving velocity so that the injected power is preserved when plotting the point in the parameter plane used here. For $\gamma=0.01\,\mathrm{pN\,s}/\mu\mathrm{m}$, this gives $v\simeq 15.8\,\mu\mathrm{m/s}$ at $k=70\,\mathrm{pN}/\mu\mathrm{m}$, corresponding to an injected power $P\simeq 2.5\,\mathrm{aW}$. The resulting operating point, indicated by a yellow star in Fig.~\ref{fig:design_chart}, lies in a regime of strong force anti-alignment, with $r_{\mathrm{ss}}\simeq -0.96$.

 { 
The RBC experiments in Ref.~\cite{DiTerlizzi2024} are described using an active-particle model with coupled membrane and cortex variables and a time-correlated active force \footnote{This mapping should be interpreted as a placement of an experimentally realized constant-drag protocol within the force-geometric chart, not as a direct model of the entire results of the RBC membrane-flickering experiment in Ref.~\cite{DiTerlizzi2024}.}. In the notation of our framework, this corresponds to an external force containing both trapping force and an active contribution, $ \vec F_{\rm ext} =\vec F_{\rm trap}+\vec F_{\rm active}$. The active force in that model is an Ornstein--Uhlenbeck process and therefore has zero steady-state mean, but it drives persistent fluctuations around the static trapped state. The RBC setting is distinct from the constant-drag protocol marked in Fig.~\ref{fig:design_chart}, in the sense that the latter has nonzero mean transport, whereas the former is a confined active-fluctuation problem without transport. We use the marked point to show that experimentally accessible dragged-particle parameters can lie close to the anti-aligned regime in the force-geometric representation. In other words,  this operating point indicates that even when fluctuations are dominant, entropy production can be small. This aligns with the experimental observation that dissipation can decrease for higher fluctuations. The RBC experiments motivate extending the same geometric framework to active-fluctuation systems.}

\subsection{Sinusoidal driving: force geometry and microrheological operation}

In optical-tweezer microrheology and related experiments, the center of a harmonic trap
is often modulated sinusoidally with frequency $\omega$:
$\lambda(t)=A\cos(\omega t)$, to probe frequency-dependent mechanical response
\cite{Kumar2021Multi,CatalaCastro2025}.
For overdamped dynamics, the mean particle position $\mu(t)$ follows the drive with a
phase lag $\phi$ determined by
$\tan\phi=\gamma\omega/k$, where $k$ is the trap stiffness and $\gamma$ the effective
friction coefficient.

For this protocol, the force--correlation coefficient, time-averaged over one driving
period $\tau=2\pi/\omega$, can be evaluated exactly (see Appendix~\ref{AppenGeneralFrameworkForMovingHarmonicTraps} for derivation and Appendix~\ref{AppenExpMapping} for mapping to the experimental data):
\begin{equation}
\langle r\rangle_\tau
:= \frac{1}{\tau}\int_0^\tau r(t)\,dt = -\frac{2}{\pi}\,
K\!\left(-\alpha \right),
\label{eq:r_exact_main}
\end{equation}
where $K(\cdot)$ denotes the complete elliptic integral of the first kind, $\alpha = Q\sin^2\phi$, and
$Q=kA^2/(k_BT)$ is a dimensionless measure of the driving amplitude relative to thermal
fluctuations.

 {The corresponding cycle-averaged entropy production rate and injected mechanical power is
\begin{align}
\langle \dot S_i\rangle_\tau
= \frac{k k_B}{2\gamma}\,\alpha, \qquad
\langle P\rangle_\tau
=
T\langle \dot S_i\rangle_\tau
=
\frac{k k_BT}{2\gamma}\,\alpha,
\end{align}
Thus, for the sinusoidally driven harmonic trap, both the energetic cost and the force correlation are governed by the same dimensionless lag parameter $\alpha=Q\sin^2\phi$. The parameter $\alpha$ has a simple interpretation: it is the squared amplitude of the mean lag measured in units of the steady-state variance.  In the fluctuation-dominated regime, $\alpha\ll1$, the elliptic-integral expression for the cycle-averaged force correlation gives
\begin{align}
\langle r\rangle_\tau
=
-1+\frac{\alpha}{4}+O(\alpha^2),
\end{align}
and hence
\begin{align}
\langle \dot S_i\rangle_\tau
\simeq
\frac{2k k_B}{\gamma}
\left(1+\langle r\rangle_\tau\right).
\end{align}}

 {These relations show how force correlation organizes energetic cost in a protocol-dependent way. In the fluctuation-dominated regime, the deterministic lag is small compared with the thermal width, so the external and information-theoretic forces are close to anti-aligned, $\langle r\rangle_\tau\to-1$, and both the cycle-averaged entropy production and injected power are small. Conversely, finite injected power requires a nonzero lag, so reducing $\alpha$ moves the system closer to anti-alignment while also lowering the injected power.  Perfect anti-alignment is reached only in the limiting case $\alpha\to0$, for example in the quasistatic limit $\phi\to0$ at fixed $Q$, or in the vanishing-amplitude limit $Q\to0$ at fixed $\phi$. At finite $\alpha$, force cancellation is incomplete and the dynamics remain irreversible. Since both $\langle r\rangle_\tau$ and $\langle P\rangle_\tau$ are functions of the same parameter $\alpha=Q\sin^2\phi$, increasing $\alpha$ simultaneously moves the system away from anti-alignment and increases the energetic cost. Thus, within this harmonic sinusoidal protocol, the geometry of force alignment directly organizes the energetic cost, as contours of constant force correlation coincide with contours of constant injected power in the $(Q,\phi)$ plane.}

Figure~\ref{fig:sinusoidal_design_chart} summarizes this geometric organization as a chart.
Dashed curves indicate contours of constant force correlation which, as noted, coincide with contours of constant time-averaged injected power.
Also shown is the experimentally relevant driving strength $Q_{\mathrm{ext}} =486.6$, fixed by the oscillation amplitude $A$, trap stiffness $k$, and thermal
scale $k_BT$, corresponding to the parameters used in Ref.~\cite{CatalaCastro2025}. Details of the parameters used to map experimental conditions onto the
theoretical framework for both driving protocols (including the rescaling
required for constant-velocity driving) are provided in Appendix~\ref{AppenExpMapping}.
 Varying the driving frequency corresponds to traversing this line across the accessible
range of phase lags. Within this representation, the fixed-$Q_{\mathrm{ext}}$ trajectory intersects contours
of relatively strong force anti-alignment ($\langle r\rangle_\tau \lesssim -0.9$) at small
phase lags.
Because the phase lag is set by the driving frequency through
$\tan\phi=\gamma\omega/k$,  the chart shows how changing frequency at fixed
\(Q_{\mathrm{ext}}\) 
jointly tune force anti-alignment and injected power.

This geometric organization provides a principled basis for navigating
trade-offs between dissipative cost and mechanical response in
microrheological operation.
From this viewpoint, dynamical regimes already exploited in microrheological
experiments  can be interpreted  as manifestations of an underlying force geometry. By resolving how forces are organized, this framework makes explicit structural features of entropy production that remain implicit in energetic analyses.

 {\section{Richness in $d>1$ dimensions}}
\label{sec:high-dimensional-angles}

\subsection{The example of a two dimensional harmonic trap} 
To illustrate the emergence of continuous force-angle structure in a two dimensional system, consider an overdamped Brownian particle in a static isotropic harmonic potential,
\begin{align}
U(x,y)=\frac{k}{2}(x^2+y^2),
\end{align}

whose center is fixed at the origin. At a given instant {during transient relaxation}, let the probability density be an isotropic Gaussian with mean displaced from the trap center by $\vec{\mu}=(a,b)$ and variance $\sigma^2$. The external and information-theoretic forces are
\begin{align}
\vec{F}_{\mathrm{ext}}(x,y)
&=
-\vec{\nabla} U
=
-kx\,\hat{\mathbf{i}}-ky\,\hat{\mathbf{j}},
\\
\vec{F}_{\mathrm{info}}(x,y)
&=\frac{k_B T}{\sigma^2}\left((x-a)\hat{\mathbf{i}} + (y-b)\hat{\mathbf{j}}\right).
\end{align}
\begin{figure}[h!]
\centering
\hspace{-0.5cm}\includegraphics[width=0.5\textwidth]{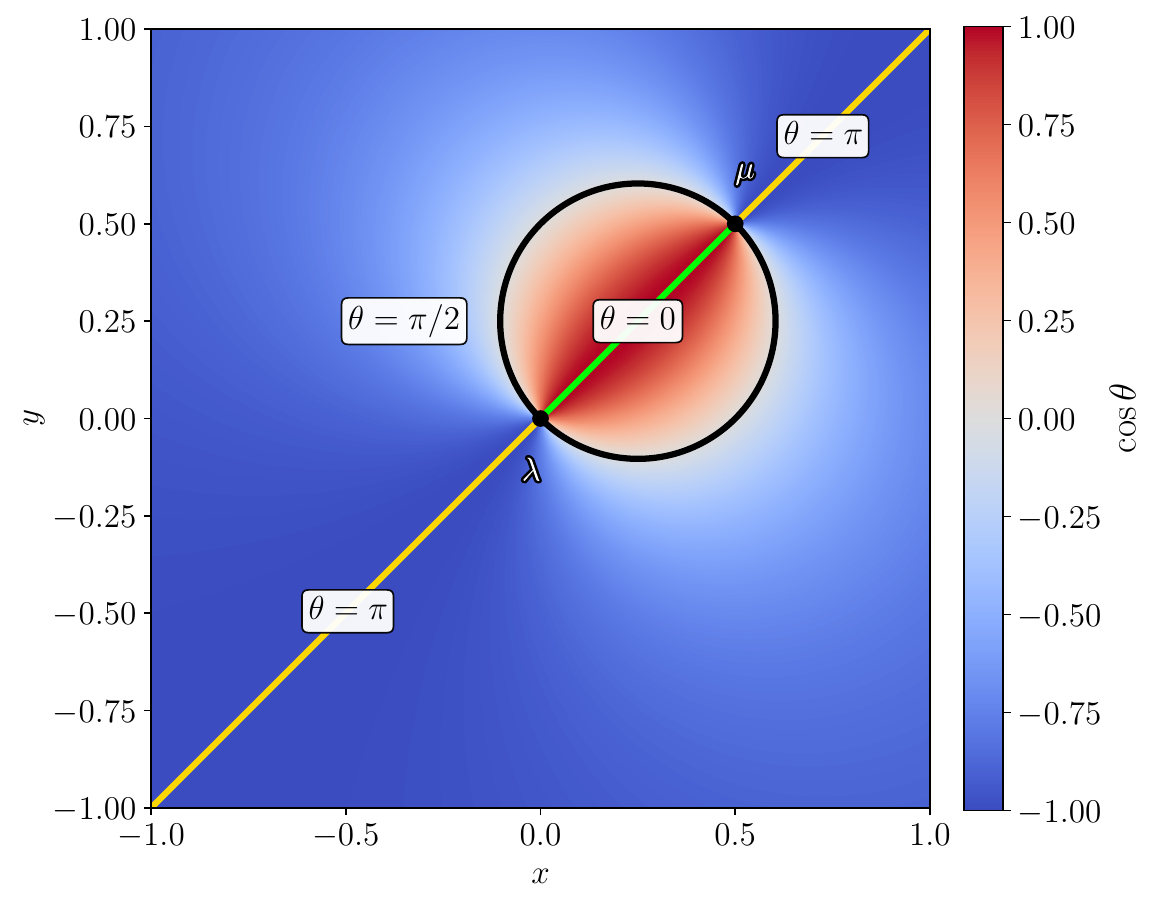}
\caption{\footnotesize{Local force-angle field in two dimensions. The color scale represents $\cos[\theta(x,y)]$, and the black contour marks $\theta=\pi/2$. Here $s$ is the line parameter in $(x,y)=s(a,b)$. The green line segment marks the aligned contour $\theta=0$ along the straight line joining $\lambda=(0,0)$ and $\mu=(a,b)$ for $0<s<1$, while the yellow half-lines mark the anti-aligned contour $\theta=\pi$ on the same line for $s<0$ and $s>1$. For the plotted example, $a=b=0.5$.}}
\label{fig:two_dim_model}
\end{figure}
The local angle between the two force fields is 
\begin{align}
\cos[\theta(x,y)]
&=
\frac{
\vec{F}_{\mathrm{ext}}(x,y)\cdot
\vec{F}_{\mathrm{info}}(x,y)
}{
\left|\vec{F}_{\mathrm{ext}}(x,y)\right|
\left|\vec{F}_{\mathrm{info}}(x,y)\right|
} \nonumber\\ 
&= 
-
\frac{
x(x-a)+y(y-b)
}{
\sqrt{x^2+y^2}
\sqrt{(x-a)^2+(y-b)^2}}.
\label{eq:2Dcostheta}
\end{align}

Unlike the one-dimensional case, the two-dimensional system exhibits a continuous distribution of local force angles across configuration space. The three force-alignment cases, together with their corresponding contours, can be summarized as: 
{\small
\begin{alignat}{5}
&\vec{F}_{\mathrm{info}} \parallel \vec{F}_{\mathrm{ext}}
&&\;\Leftrightarrow\;
&&\theta = 0
&&\;\Leftrightarrow\;
&&(x,y)=s(a,b), \quad 0<s<1,
\label{eq:theta_zero_force_relation}\\[4pt]
&\vec{F}_{\mathrm{info}} \perp \vec{F}_{\mathrm{ext}}
&&\;\Leftrightarrow\;
&&\theta = \tfrac{\pi}{2}
&&\;\Leftrightarrow\;
&&\left(x-\tfrac{a}{2}\right)^{\!2}+\left(y-\tfrac{b}{2}\right)^{\!2}=\tfrac{a^2+b^2}{4},
\label{eq:theta_halfpi_force_relation}\\[4pt]
&\vec{F}_{\mathrm{info}} \parallel -\vec{F}_{\mathrm{ext}}
&&\;\Leftrightarrow\;
&&\theta = \pi
&&\;\Leftrightarrow\;
&&(x,y)=s(a,b), \quad s<0 \text{ or } s>1.
\label{eq:theta_pi_force_relation}
\end{alignat}
}where $s\in\mathbb{R}$ parametrizes the line joining the trap center $(0,0)$ and the probability-density center $(a,b)$.

In particular, the contour along which the external and information forces are orthogonal is given by the circle appearing in the middle line of Eq.~\eqref{eq:theta_halfpi_force_relation}.
Thus, the locus of perpendicular forces forms a circle centered at $(a/2,b/2)$ with radius $\sqrt{a^2+b^2}/2$. This contour separates regions where the two forces form an acute angle,
$ 0 \le \theta < \frac{\pi}{2}$, from regions where they form an obtuse angle, $\frac{\pi}{2}<\theta\le\pi$.

The resulting angle field provides a simple geometric visualization, see Fig.~\ref{fig:two_dim_model}, of force organization in higher dimensions. While the local force geometry is described by the continuous field $\theta(x,y)$, the force-correlation coefficient $r$ compresses the underlying spatial distribution of force angles into a single scalar quantity through a $\rho$-weighted averaging.
\\

\subsection{Orthogonal force geometry in central-force systems}
The force-angle framework is not restricted to harmonic confinement. Consider a Brownian particle subject to an arbitrary central potential $U(r)$, for which the external force is purely radial,
\begin{align}
\vec{F}_{\mathrm{ext}}
=
-\frac{dU}{dr}\,\hat{r}.
\end{align}

For a general probability density $\rho(r,\phi,t)$, the information-theoretic force is
\begin{align}
\vec{F}_{\mathrm{info}}
=
-k_B T
\left(
\frac{\partial \ln \rho}{\partial r}\,\hat{r}
+
\frac{1}{r}
\frac{\partial \ln \rho}{\partial \phi}\,\hat{\boldsymbol{\phi}}
\right).
\end{align}
Since the external force is purely radial, orthogonal force geometry arises whenever $\partial_r \rho = 0$, for which $\vec{F}_{\mathrm{info}}$ becomes purely tangential and $\theta=\pi/2$. 

A particle confined near a ring of radius $R$ represents an idealized limiting case {of a transient state} in which radial variations of the probability density become negligible, $\partial_r \rho = 0$, while the external force remains purely radial. The information-theoretic force therefore reduces to
\begin{align}
\vec{F}_{\mathrm{info}}
=
-\frac{k_B T}{R}
\frac{\partial \ln \rho}{\partial \phi}\,\hat{\boldsymbol{\phi}},
\end{align}
which is purely tangential for any admissible probability density along the ring. Since $\hat{r}\cdot\hat{\boldsymbol\phi}=0$, the two forces remain orthogonal {on this contour at that instant},
\begin{align}
\vec{F}_{\mathrm{ext}}\cdot\vec{F}_{\mathrm{info}}=0,
\qquad
\theta=\frac{\pi}{2}.
\end{align}
{On this contour, a nonzero radial external force can still drive a radial probability current and contribute to the local entropy-production density.}
{In the idealized near-ring limit, orthogonality extends throughout the populated region, yielding}
\begin{align}
r(t)=0,
\end{align}
{and the entropy production rate becomes}
\begin{align}
\dot{S}_{i}
=
\frac{D}{k_B T^2}
\left(
\left\langle
|\vec{F}_{\mathrm{info}}|^2
\right\rangle
+
\left\langle
|\vec{F}_{\mathrm{ext}}|^2
\right\rangle
\right),
\end{align}
where the correlation contribution vanishes identically due to the orthogonality of the force vectors. This example demonstrates that the force-geometry framework extends beyond harmonic systems and accommodates qualitatively different geometric organizations of entropy production. In contrast to the thermodynamic stall condition $(\theta=\pi,\, r=-1)$, diffusion on a ring realizes an orthogonal geometry $(\theta=\pi/2,\, r=0)$, for which the external and information-theoretic forces contribute only additively to the entropy production. More generally, in central-force systems, local orthogonal force geometry arises on contours satisfying $\partial_r \rho = 0$, where the information-theoretic force becomes purely tangential while the external force remains radial.  The two-dimensional harmonic trap discussed previously provides an example, exhibiting a $\theta=\pi/2$ contour separating locally aligned and anti-aligned regions of configuration space.

\section{Relation to existing thermodynamic bounds} 
\label{sec:existing-thermodynamic-bounds}
\label{SecRelationToExistingBounds}
Thermodynamic uncertainty relations (TURs) bound the precision of a prescribed
current in terms of the entropy production rate, placing fundamental limits on
fluctuations at a given energetic cost
\cite{Barato2015Thermodynamic,Gingrich2016Dissipation}.
These results play an essential role in characterizing the reliability of
nonequilibrium processes, but they address a distinct aspect of
nonequilibrium behavior to the present framework.
TURs assume a fixed entropy production budget and ask how precisely a current
can be realized; by contrast, the force-correlation framework focuses on how
entropy production is structured through force organization.
In this sense, TURs constrain performance \textit{given} entropy production,
whereas force geometry clarifies a mechanism by which entropy production itself
is generated. 
{We also note that at stall, where the mean displacement current vanishes, the TUR for that current becomes
trivial, while the entropy production rate can remain finite and is then determined entirely by fluctuations of
the net force.}
Together, these perspectives provide a more complete thermodynamic description,
combining scalar bounds on precision with a geometric view of force alignment.

A parallel standpoint is provided by Fisher-information--based speed limits discussed in Ref.~\cite{Nicholson2020TimeInformation}.
These bounds constrain the minimal time required for a system to evolve between
probability distributions by relating entropy production to the statistical
distinguishability of states, and they apply even to underdamped dynamics.
While powerful, such limits remain agnostic to the organization of forces:
the same state-space displacement can be realized through distinct force
configurations with different entropy production.
By contrast, the force-correlation framework resolves the vectorial organization
of thermodynamic forces, clarifying how alignment between driving and information-theoretic
forces shapes entropy production.

Related geometric views on entropy production have also been developed,
e.g., through decompositions based on information geometry and
thermodynamic length~\cite{Ito2020Geometric}.
These approaches operate at the level of probability distributions and currents,
providing valuable bounds on entropy production in finite-time processes.
Our analysis complements these perspectives by explicitly resolving
force alignment within the thermodynamic description.

In totality, these perspectives illuminate distinct structural aspects of
nonequilibrium dynamics.
Thermodynamic uncertainty relations constrain the precision of currents at fixed
entropy production, Fisher-information speed limits bound the minimal time required for
state evolution, and force geometry clarifies how entropy production is generated through
the geometric organization of competing forces.
Integrating these viewpoints places force geometry within the broader
framework of nonequilibrium constraints, alongside statistical bounds
on speed and precision.

\section{Summary and Outlook}
\label{sec:summary-outlook}
In this work, we identify force alignment as a structural determinant
of irreversibility in overdamped nonequilibrium dynamics.
By decomposing the thermodynamic force into deterministic and information-theoretic components,
we show that entropy production depends not only on force magnitudes, but also on their
geometric organization. We quantify this organization through a force correlation coefficient.
This perspective complements scalar current-based bounds on thermodynamic observables by restoring the vectorial structure of thermodynamic forces.

Throughout, we focus on overdamped dynamics where the thermodynamic force admits
a decomposition into external and information-theoretic contributions; extending the framework
to systems with additional force components or active noise would require a
corresponding extension of the force-space description.

In overdamped systems, thermodynamic stall follows from the proportionality between mean velocity and mean net force $(d_t\langle \vec{x} \rangle \propto \langle \vec{\mathcal{F}}_{\mathrm{net}} \rangle)$. Perfect anti-alignment cancels the average force and eliminates mean transport, while entropy production remains finite, Fig.~\ref{box:geometric_hierarchy}. In this case, entropy production is zero only if forces cancel pointwise in space and time. This separation between transport suppression and true reversibility highlights
that geometric anti-alignment represents a distinct nonequilibrium condition.
Extending this perspective beyond overdamped dynamics (see Appendix~\ref{Appen_underdamped}), where additional 
degrees of freedom such as momentum are present, suggests a richer relationship
between force geometry and transport.
In underdamped dynamics, momentum acts as a thermodynamic buffer that partially decouples force geometry from transport. This allows for the coexistence of strong positional anti-alignment and directed motion -- a dynamical state where transport persists even as positional force cancellation suppresses instantaneous entropy production. Such dynamics identify potential ``coasting'' regimes that scalar descriptions, lacking access to the underlying vectorial organization, would otherwise conflate with simple dissipative dragging in the presence of friction. Recent experiments on underdamped micro-mechanical oscillators~\cite{Ludovic2021}
provide a natural setting to explore how force organization, transport,
and irreversibility are jointly controlled.

A complementary direction is suggested by recent experiments in living systems. In Ref.~\cite{DiTerlizzi2024}, steady-state entropy production was inferred from high-resolution flickering experiments in human red blood cells, including cell stretching and contour fluctuations analyzed as a proxy for metabolic activity. From the present perspective, these observations invite a geometric reinterpretation of low-dissipation operation in living matter. Sustained activity far from equilibrium may reflect the dynamical orchestration of internally generated and externally imposed forces. In overdamped environments, near-cancellation between internally generated forces and external loads can reduce the fluctuating component of the net thermodynamic force, thereby lowering the excess entropy production associated with a prescribed mean net force.  {In this view, force anti-alignment provides a potential mechanism for optimizing entropy production in overdamped systems: once $\langle\vec{\mathcal{F}}_{\rm net}\rangle$ is fixed, including the stall case $\langle\vec{\mathcal{F}}_{\rm net}\rangle=0$, the design principle becomes $\min\,{\rm Var}(\vec{\mathcal{F}}_{\rm net})$ subject to fixed $\langle\vec{\mathcal{F}}_{\rm net}\rangle$.}

Beyond inertia, spatial dimensionality introduces additional geometric freedom. Our geometric view extends directly to higher dimensions, where forces become vector fields and scalar products are replaced by inner products, without requiring additional assumptions. While force geometry in one-dimensional confinement is necessarily binary (aligned or anti-aligned), higher dimensions permit continuous angular reorganization of competing forces, allowing partial cancellation and transport to coexist. Together with recent optimal-transport bounds on finite-time entropy production~\cite{Oikawa2025}, these considerations suggest that force geometry offers a unifying structural perspective and a potential framework for organizing and regulating nonequilibrium operation across driven systems.

\appendix 
\setcounter{secnumdepth}{1}
\renewcommand{\thesection}{\Alph{section}}
\renewcommand{\theHsection}{\thesection}

\section{Deriving the entropy production rate from the informatio-theoretic force}
\label{AppenDervEntrp}

According to standard formulations of stochastic thermodynamics
\cite{Seifert2012Stochastic},
the time derivative of the system entropy and the entropy flow to the environment
can be expressed in terms of the time-dependent probability density $\rho(\vec{x},t)$ as
\begin{align}
\text{(i)}\quad
\dot{S}(t)
&= -k_B \int_{-\infty}^{\infty}
\dot{\rho}(\vec{x},t)\,\ln\rho(\vec{x},t)\,d\vec{x}, \\
\text{(ii)}\quad
\dot{S}_e(t)
&= \frac{1}{T} \int_{-\infty}^{\infty}
\dot{\rho}(\vec{x},t)\,U(\vec{x},t)\,d\vec{x},
\label{EqA1A2}
\end{align}
where the notation $\int d\vec{x}$ denotes a volume integral over all the degrees of freedom. The entropy production rate is 
$\dot{S}_i(t) \equiv \dot{S}(t) - \dot{S}_e(t)$. From this relation, using  the Fokker--Planck Eq.~\eqref{EqFP} for $\dot{\rho}(\vec{x},t)$, together with Eq.~\eqref{EqA1A2} and the boundary conditions $\lim_{x_n\rightarrow\pm\infty}\rho\rightarrow0$, $\lim_{x_n\rightarrow\pm\infty}{\partial_{x_n}\rho}\rightarrow0$ for every $n\in[1,d]$, we find:
\begin{align}
\dot S_i
&=
k_B D
\int d\vec{x}\,
\rho(\vec{x},t)
\left|
\vec{\nabla}\ln\rho(\vec{x},t)
+
\frac{\vec{\nabla} U(\vec{x},t)}
{k_B T}
\right|^2.
\end{align}
Here, we identify $\vec{F}_\text{ext}(\vec{x},t)=-\vec{\nabla}U(\vec{x},t)$ and, defining the information-theoretic force, $\vec{F}_{\mathrm{info}}(\vec{x},t) = -k_B T \vec{\nabla} \ln \rho(\vec{x},t)$, we obtain $\dot S_i(t)$ in Eq.~\eqref{EqNetForce}. 
\\ 

\begin{table*}[t]
    \centering
    \renewcommand{\arraystretch}{1.75}
    \begin{tabular}{|l|c|c|}
        \hline
        \textbf{Quantity}\rule{0pt}{5.5ex} & 
        \textbf{Current Representation}\rule{0pt}{5.5ex} & 
        \textbf{Force Representation}\rule{0pt}{5.5ex} \\
        \hline
        Local mean velocity $\vec{v}(\vec{x},t)$ & 
        $\displaystyle \frac{\vec{J}(\vec{x},t)}{\rho(\vec{x},t)}$ & 
        $\displaystyle \mu\,\vec{\mathcal{F}}_{\mathrm{net}}(\vec{x},t)$ \\
        \hline
        
        Mean transport velocity $\frac{d}{dt}\langle\vec{x}\rangle$ & 
        $\displaystyle \int d\vec{x}\;\vec{J}(\vec{x},t)$ & 
        $\displaystyle \mu \left\langle \vec{\mathcal{F}}_{\mathrm{net}} \right\rangle$ \\
        \hline
        
        Entropy production rate $\dot{S}_i\rule{0pt}{5.5ex}$ & 
        $\begin{aligned}
            &k_B\int d\vec{x}\,\frac{|\vec{J}(\vec{x},t)|^2}{D\rho(\vec{x},t)} = \frac{k_B}{D}\left\langle |\vec{v}|^2 \right\rangle
        \end{aligned}$ & 
        $\displaystyle \frac{D}{k_B T^2}
        \left\langle |\vec{\mathcal{F}}_{\mathrm{net}}|^2 \right\rangle$ \\
        \hline
    \end{tabular}
    \caption{Comparison of physical quantities for overdamped stochastic dynamics in the current and force representations. The angular brackets denote averages with respect to the instantaneous probability density $\langle A \rangle=\int d\vec{x}\,\rho(\vec{x},t)\,A(\vec{x},t)$, $\vec{\mathcal{F}}_{\rm net} = \vec{F}_{\rm ext} + \vec{F}_{\rm info}$, where  $\vec{F}_{\rm info}= -k_BT\vec\nabla\ln\rho$, and the Einstein-Smoluchowski relation $D = \mu k_BT$ is used. The local mean velocity in the table coincides with the Stratonovich average of the displacement rate conditioned on a given position~\cite{Seifert2012Stochastic}.}
    \label{tab:velocity-force-representations}
\end{table*}

Note that using the information-theoretic force we can express also the  entropy production rate in a manner similar to the total entropy growth and  flow  rates presented in~\cite{Nicholson2020TimeInformation,DechantSasa2021PhysRevX11.041061},  through the surprisal $I(\vec{x},t)=-\log(\rho(\vec{x},t))$:
\begin{align}
\dot S(t)
&= -k_B\left\langle \dot I(\vec{x},t)\, I(\vec{x},t)\right\rangle, \nonumber\\
\dot S_e(t)
&= -\frac{1}{T}\left\langle U(\vec{x},t)\dot I(\vec{x},t)\right\rangle\nonumber\\ 
\dot{S}_i(t)
&= k_B D\,
\bigl\langle |\vec{\nabla}\left[\tilde{U}(\vec{x},t)- I(\vec{x},t)\right]|^2 \bigr\rangle, 
\end{align} 
where $\tilde{U}(\vec{x},t)=U(\vec{x},t)/k_B T$.
With the definitions $\vec{F}_\text{ext}=-\vec{\nabla}U$ and $\vec{F}_{\mathrm{info}}=-k_B T\,\vec{\nabla}\ln\rho$, the square in the preceding expression is equivalently proportional to $\left|\vec{F}_\text{ext}+\vec{F}_{\mathrm{info}}\right|^2$. Also, see Table~\ref{tab:velocity-force-representations}.

 {\section{Force correlation coefficient and thermodynamic stall }}
\label{Appen_force_coeff_stall}
In the  force-geometric setting, $r(t)=-1$ is a sufficient condition for stall (zero mean transport). To see this, we first recall that for overdamped dynamics with constant mobility $\mu$,
$d_t\langle \vec x\rangle=
\mu \langle \vec {{\mathcal{F}}}_{\rm net}\rangle$.

For thermodynamic stall,  $\langle\vec {\mathcal{F}}_{\rm net}\rangle$ must vanish. We next show that $\langle\vec {\mathcal{F}}_{\rm net}\rangle = 0$ corresponds to $r = -1$.

The condition $r(t)=-1$ corresponds to situation in which the external and information-theoretic forces are locally anti-aligned:
\begin{align}
\vec F_{\rm ext}(\vec x,t)=
-c\vec F_{\rm info}(\vec x,t),
\qquad c>0,
\end{align}
which implies that
\begin{align}
\vec{F}_{\rm net} = \vec F_{\rm ext}+\vec F_{\rm info} = (1-c) \vec F_{\rm info},    \quad (r = -1).
\end{align}

At $r = -1$, there are two cases to consider:
\begin{enumerate}
\item $\vec{F}_{\rm net} = 0$ (for $c = 1$) corresponds to the case when $\vec F_{\rm ext}(\vec x,t)$ and $\vec F_{\rm info}(\vec x,t)$ cancel each other pointwise.  This is the equilibrium condition, $\dot S_i = 0$.
\item In the case when $c \neq 1$,  $\vec{F}_{\rm net} $ does not vanish identically at all points in space. However, the mean $\langle\vec{F}_{\rm net}\rangle$ is still zero. This is because the information-theoretic force $\vec F_{\rm info}$ has zero average under the standard boundary conditions:
\begin{align}
\langle \vec F_{\rm info}\rangle
&=
-k_B T\int \rho \nabla\ln\rho\,d\vec x  \nonumber\\
&=
-k_B T\int \nabla\rho\,d\vec x
=
0 .
\end{align} 
Thus, at $r = -1$,  $\langle\vec{F}_{\rm net}\rangle = (1-c)\langle \vec F_{\rm info}\rangle = 0$. 
\end{enumerate}
Therefore, unless the local net force is zero pointwise, the entropy production rate remains finite:
\begin{align}
\dot S_i
&=
\frac{D}{k_B T^2}
\left\langle
|\vec {\mathcal{F}}_{\rm net}|^2
\right\rangle \nonumber\\
&= \frac{D}{k_B T^2}
\text{Var}\,\vec{\mathcal{F}}_{\rm net} \neq 0. \quad (r = -1, \vec{\mathcal{F}}_{\rm net} \neq 0)
\end{align}
Thus $r = -1$ corresponds to the thermodynamic stall with zero mean transport but generally finite entropy production. The converse implication does not hold in general: stall does not by itself enforce $r=-1$.

\section{Integrated Bounds for Finite-Time Processes} 
\label{AppenIntegratedBounds}

\paragraph{From Instantaneous to Integrated Constraints.} 
While instantaneous bounds provide moment-by-moment constraints on entropy production rate, practical
experiments and engineered protocols operate over finite durations.
Here, assuming isotropic overdamped diffusion,  we derive integrated bounds that constrain the total entropy production over
an arbitrary protocol time~$\tau$.
\\ 

\paragraph{Derivation of the Integrated Bound.}
Starting from the instantaneous inequality, Eq.~\eqref{EqInstanteneousInequality}, we integrate the
entropy production rate over the protocol duration~$\tau$,
\begin{align}
\hspace{-1em}\Sigma_\tau &= \int_0^\tau \dot S_i(t)\,dt \nonumber\\
&\ge
\frac{D}{k_B T^2}
\int_0^\tau
\left(
\sqrt{\langle |\vec{F}_{\text{ext}}|^2\rangle_t}
-
\sqrt{\langle |\vec{F}_{\text{info}}|^2\rangle_t}
\right)^2
dt ,
\end{align}
where $\Sigma_\tau$ is the total entropy production.

Defining
\begin{equation}
g(t)
=
\sqrt{\langle |\vec{F}_{\text{ext}}|^2\rangle_t}
-
\sqrt{\langle |\vec{F}_{\text{info}}|^2\rangle_t},
\end{equation}
the bound becomes
\begin{equation}
\Sigma_\tau
\ge
\frac{D}{k_B T^2}
\int_0^\tau g^2(t)\,dt .
\end{equation}
Applying Jensen's inequality to the convex function \(x\mapsto x^2\), with
\(g(t)\) as its argument, gives
\begin{equation}
\frac{1}{\tau}\int_0^\tau g^2(t)\,dt
\ge
\left[
\frac{1}{\tau}\int_0^\tau g(t)\,dt
\right]^2 .
\end{equation}
Therefore,
\begin{equation}
\Sigma_\tau
\ge
\frac{D\tau}{k_B T^2}
\left(
\overline{\sqrt{\langle |\vec{F}_{\text{ext}}|^2\rangle}}
-
\overline{\sqrt{\langle |\vec{F}_{\text{info}}|^2\rangle}}
\right)^2 ,
\end{equation}
with time averages defined as
\begin{align}
\overline{\sqrt{\langle |\vec{F}_{\text{ext}}|^2\rangle}}
&=
\frac{1}{\tau}
\int_0^\tau
\sqrt{\langle |\vec{F}_{\text{ext}}|^2\rangle_t}\,dt , \nonumber\\
\overline{\sqrt{\langle |\vec{F}_{\text{info}}|^2\rangle}}
&=
\frac{1}{\tau}
\int_0^\tau
\sqrt{\langle |\vec{F}_{\text{info}}|^2\rangle_t}\,dt .
\end{align}

\paragraph{Scope and applicability.}
The integrated bounds derived here rely only on the Markovian overdamped
Fokker--Planck description and therefore apply far from equilibrium, without
invoking linear-response or quasistatic assumptions.
They hold for arbitrary protocol durations $\tau>0$, from instantaneous limits
to finite-time processes, and are independent of the specific initial state preparation.
Moreover, the bounds are valid for general time-dependent driving protocols
$U(\vec{x},t)$ and do not assume any specific functional form of the driving.

\paragraph{Physical Interpretation.}
The integrated bound shows that the total entropy production over a finite-time protocol is
constrained by the difference between the time-averaged root mean squared magnitudes of the external and information forces.
The instantaneous force-correlation structure determines how closely this bound can be
approached over the protocol duration.
For closely matched force magnitudes, protocols that maintain favorable force anti-alignment throughout the protocol
accumulate less excess entropy production, whereas protocols with poor force alignment incur
larger geometric waste gaps, even when leading to identical macroscopic outcomes.
In the overdamped setting considered here, favorable force anti-alignment corresponds
to proximity to stall rather than true reversibility, since pointwise force cancellation is not imposed.
\\

\paragraph{Relation to Current-Based Integrated Bounds.}
Unlike thermodynamic uncertainty relations and related integrated bounds, which provide
scalar constraints on total entropy production or current fluctuations, the present
inequality retains explicit information about the time-resolved force structure along the protocol.
This decomposition makes it possible to identify which segments of a protocol contribute most strongly to
entropy production, thereby enabling a localized attribution of entropy production rather than a purely global
estimation.
Correlation-aware thermodynamic bounds have been studied in other contexts, where incorporating correlations
between observables can tighten traditional scalar bounds such as the thermodynamic uncertainty
relation~\cite{DechantSasa2021PhysRevX11.041061}, but these frameworks operate at the level of stochastic observables,
rather than directly resolving the geometric organization the two forces considered here.
\\

\paragraph{Implications for Work Bounds.}
For isothermal processes coupled to a single thermal reservoir and using the Sekimoto
definition of stochastic work (with Stratonovich interpretation), the total excess work
satisfies
\begin{equation}
W_\tau - \Delta A = T\,\Sigma_\tau ,
\end{equation}
where $\Delta A$ represents Helmholtz free energy difference. 

This implies the bound
\begin{equation}
W_\tau \ge
\Delta A
+
\frac{D\tau}{k_B T}
\left(
\overline{
\sqrt{
\left\langle
|\vec{F}_{\text{ext}}|^2
\right\rangle
}
}
-
\overline{
\sqrt{
\left\langle
|\vec{F}_{\text{info}}|^2
\right\rangle
}
}
\right)^2 .
\end{equation}
This inequality establishes a minimum energetic cost required to execute a protocol of
duration~$\tau$, independent of the detailed time dependence of the driving.
Unlike quadratic-response bounds or speed limits, this constraint explicitly links
energetic expenditure to the geometric mismatch between external and information forces.

\paragraph{Experimental Utility.}

The integrated bound provides experimentally actionable diagnostics for interpreting protocol performance.
Reducing excess entropy production does not require suppressing dynamics or extending protocol durations,
but rather maintaining close correspondence between
$\sqrt{\langle |\vec{F}_{\text{ext}}|^2\rangle_t}$ and
$\sqrt{\langle |\vec{F}_{\text{info}}|^2\rangle_t}$ throughout the protocol, so that their time-averaged mismatch stays small.
This time-resolved viewpoint enables systematic identification of geometrically inefficient segments in complex driving protocols, clarifying where excess entropy production accumulates.

\section{General Framework for Moving Harmonic Traps}
\label{AppenGeneralFrameworkForMovingHarmonicTraps}

\paragraph{System Description and Dynamics.}
We consider a Brownian particle immersed in a viscous fluid at temperature $T$, subject to a harmonic trapping potential whose centre $\lambda(t)$ can be externally controlled~\cite{Risken1989Fokker,vanKampen2007Stochastic}. The particle experiences deterministic restoring forces from the trap and random thermal fluctuations from the surrounding medium~\cite{Sekimoto2010Stochastic}.

The potential energy is
\begin{equation}
U(x,t) = \frac{1}{2}k\big[x - \lambda(t)\big]^2,
\end{equation}
The corresponding deterministic force is
\begin{equation}
F_{\text{ext}}(x,t) = -\nabla U = -k\big[x - \lambda(t)\big],
\end{equation}
which always points toward the instantaneous trap centre $\lambda(t)$.

In the overdamped regime, where inertia is negligible compared to viscous
friction, the dynamics are governed by the Langevin equation
\begin{equation}
\gamma \frac{dx}{dt}
= -k\big[x - \lambda(t)\big] + \sqrt{2\gamma k_B T}\,\xi(t),
\label{eq:langevin_general}
\end{equation}
where $\gamma$ is the drag coefficient, $k_B$ is Boltzmann's constant, and
$\xi(t)$ is Gaussian white noise with
$\langle\xi(t)\rangle=0$ and
$\langle\xi(t)\xi(t')\rangle=\delta(t-t')$.

It is convenient to introduce the relaxation rate and diffusion constant,
\begin{align}
\Gamma &= \frac{k}{\gamma}, \quad D = \frac{k_B T}{\gamma}.
\end{align}
In terms of these parameters, the Langevin equation becomes
\begin{equation}
\frac{dx}{dt} = -\Gamma x + \Gamma \lambda(t) + \sqrt{2D}\,\xi(t).
\label{eq:langevin_simple}
\end{equation}
Because this equation is linear, a Gaussian initial distribution remains
Gaussian for all times.

\paragraph{Lag and Nonequilibrium Driving.}
A central quantity characterizing nonequilibrium driving is the \emph{lag}
\begin{equation}
\Delta(t) = \mu(t) - \lambda(t),
\end{equation}
defined as the displacement between the mean particle position $\mu(t)$ and
the trap centre $\lambda(t)$. The lag quantifies the system's inability to
instantaneously follow the external driving protocol.

When $\Delta(t)=0$, the distribution is centred at the trap minimum and, for
overdamped dynamics, the system remains in instantaneous equilibrium,
corresponding to quasi-static or stall-like motion. When $\Delta(t)\neq0$, the trap moves faster than the
relaxation dynamics, generating irreversible entropy production. The magnitude of
$|\Delta|$ relative to the thermal width $\sigma$ controls the degree of
nonequilibrium driving.

The mean restoring force exerted by the trap is
\begin{equation}
\langle F_{\text{ext}} \rangle(t)
= -k\langle x-\lambda\rangle
= -k\Delta(t),
\end{equation}
and the mean particle velocity is
\begin{equation}
\frac{d}{dt}\langle x \rangle = -\frac{k}{\gamma}\Delta(t).
\end{equation}
For forward dragging with $\lambda(t)=vt$ and $v>0$, the lag $\Delta<0$ and the
mean velocity is positive, indicating directed transport.

\subsection{Constant Dragging Protocol}

\label{AppenConstantDragProtocol}
\paragraph{Steady-State Solution.}
For constant-velocity dragging, $\lambda(t)=vt$, the system reaches a
nonequilibrium steady state. Solving the mean dynamics yields
\begin{equation}
\mu(t) = vt - \frac{\gamma v}{k},
\label{eq:mu_const}
\end{equation}
while the positional variance attains its equilibrium value
\begin{equation}
\sigma^2 = \langle (x-\mu)^2 \rangle = \frac{k_B T}{k},
\label{eq:sigma_const}
\end{equation}
independent of the dragging velocity. The steady-state lag is therefore
\begin{equation}
\Delta = \mu - \lambda = -\frac{\gamma v}{k},
\end{equation}
with the negative sign indicating that the distribution mean trails the trap
centre.
\\ 

\paragraph{Force Correlation Coefficient.}
The external force exerted by the moving trap is
\begin{equation}
F_{\text{ext}}(x,t) = -k(x - vt),
\end{equation}
while the information-theoretic force associated with the
probability density $\rho(x,t)$ is
\begin{equation}
F_{\text{info}}(x,t)
= -k_B T\,\frac{\partial \ln \rho}{\partial x}
= k(x-\mu),
\end{equation}
where the second equality follows from the Gaussian form of $\rho(x,t)$.

Using Gaussian statistics, the force moments are
\begin{align}
\langle F_{\text{ext}}^2 \rangle
&= k^2\big(\sigma^2 + \Delta^2\big), \\
\langle F_{\text{info}}^2 \rangle
&= k^2\sigma^2, \\
\langle F_{\text{ext}}F_{\text{info}} \rangle
&= -k^2\sigma^2.
\end{align}
The steady-state force correlation coefficient defined in the main text is therefore
\begin{equation}
r_{\mathrm{ss}}
= \frac{\langle F_{\text{ext}}F_{\text{info}} \rangle}
{\sqrt{\langle F_{\text{ext}}^2 \rangle
        \langle F_{\text{info}}^2 \rangle}}
= -\frac{1}{\sqrt{1+\Delta_{\mathrm{ss}}^2/\sigma_{\mathrm{ss}}^2}}.
\end{equation}
For a harmonic trap at finite temperature, the correlation satisfies
$-1 \le r_{\mathrm{ss}} < 0$. The strict negativity of $r_{\mathrm{ss}}$ reflects the one-dimensional harmonic geometry, in
which external and information-theoretic  forces can only align or anti-align.

Substituting the steady-state expressions
$\Delta=-\gamma v/k$ and $\sigma^2=k_B T/k$ yields
\begin{equation}
r_{\mathrm{ss}}
= -\frac{1}{\sqrt{1+\dfrac{\gamma^2 v^2}{k k_B T}}}.
\label{eq:r_exact_constAppenD}
\end{equation}
For fixed $r_{\mathrm{ss}}$, this implies the scaling relation
\begin{equation}
k \propto v^2 \qquad \text{for constant } r_{\mathrm{ss}}.
\label{eq:scaling_k_v2AppenD}
\end{equation} 

\paragraph{Injected power under constant dragging.}
For a time-dependent potential, the instantaneous injected mechanical power is
defined in the standard stochastic-thermodynamic sense as
\begin{equation}
P(t) = \left\langle \frac{\partial U(x,t)}{\partial t} \right\rangle .
\end{equation}
For a harmonically trapped particle with centre $\lambda(t)=vt$, this yields
\begin{equation}
P(t)
= -k\langle x(t)-\lambda(t)\rangle\,\dot{\lambda}(t)
= -k\,\Delta(t)\,v .
\end{equation}
In the nonequilibrium steady state, the lag $\Delta=-\gamma v/k$ is constant,
so that the time-averaged injected power becomes
\begin{equation}
\langle P\rangle
= \gamma v^2 .
\end{equation}
This expression coincides with the steady-state dissipation rate,
as expected for overdamped dynamics in contact with a single thermal reservoir.

\paragraph{Entropy Production Rate and Limiting Regimes.}
The net thermodynamic force is
\begin{align}
\vec F_{\rm net}(x,t)
&=\vec F_{\rm ext}(x,t)+\vec F_{\rm info}(x,t) \nonumber\\
&=-k(x-vt)+k(x-\mu)
=-k\Delta(t) .
\end{align}
Therefore, the steady-state entropy production rate is
\begin{align}
\dot S_i
&=
\frac{D}{k_B T^2}
\left\langle |\vec F_{\rm net}|^2 \right\rangle
\nonumber\\
&=
\frac{Dk^2}{k_B T^2}\Delta^2
=
\frac{k^2\Delta^2}{\gamma T}
=
\frac{\gamma v^2}{T}.
\end{align}
Using
\begin{equation}
r_{\mathrm{ss}}
=
-\frac{1}{\sqrt{1+\Delta_{\mathrm{ss}}^2/\sigma_{\mathrm{ss}}^2}},
\qquad
\sigma_{\mathrm{ss}}^2=\frac{k_B T}{k},
\end{equation}
the same result can also be written as
\begin{equation}
\dot S_i
=
\frac{k k_B}{\gamma}
\left(
\frac{1}{r_{\mathrm{ss}}^2}-1
\right).
\end{equation}

In the fluctuation-dominated or weak-driving regime,
\begin{equation}
\gamma v \ll \sqrt{k k_B T},
\qquad
|\Delta_{\mathrm{ss}}|\ll \sigma_{\mathrm{ss}},
\end{equation}
the forces are nearly anti-aligned, $r_{\mathrm{ss}}\simeq -1$, and
the entropy production is small, $\dot S_i\propto v^2$.

By contrast, in the drive-dominated regime,
\begin{equation}
\gamma v \gg \sqrt{k k_B T},
\qquad
|\Delta{\mathrm{ss}}|\gg \sigma{\mathrm{ss}},
\end{equation}
the global force correlation weakens toward $r_{\mathrm{ss}}\to 0^-$.
The entropy production remains $\dot S_i=\gamma v^2/T$, but it is now large
because the driving force scale dominates the thermal scale.

\subsection{Sinusoidal Driving Protocol}

\label{AppenSinusoidal}
\paragraph{Periodic Steady State.}
For sinusoidal driving, the trap centre oscillates as
\begin{equation}
\lambda(t)=A\cos(\omega t),
\end{equation}
where $A>0$ is the amplitude and $\omega$ the angular frequency.
The mean position obeys
\begin{equation}
\dot{\mu}(t)+\Gamma\mu(t)=\Gamma A\cos(\omega t),
\end{equation}
whose periodic steady-state solution is
\begin{equation}
\mu(t)=A R\cos(\omega t-\phi),
\end{equation}
with
\begin{equation}
R=\frac{1}{\sqrt{1+(\omega/\Gamma)^2}}
=\frac{1}{\sqrt{1+(\gamma\omega/k)^2}},
\quad
\tan\phi=\frac{\omega}{\Gamma}=\frac{\gamma\omega}{k},
\end{equation}
where $\phi\in[0,\pi/2)$ is the phase lag and $R\le1$ the amplitude reduction factor.
\\

\paragraph{Instantaneous Correlation Coefficient.}
The instantaneous lag between the distribution mean and the trap centre is
\begin{align}
\Delta(t) &=\mu(t)-\lambda(t)=A\!\left[R\cos(\omega t-\phi)-\cos(\omega t)\right]  \nonumber\\
&=A\sin\phi\,\sin(\omega t-\phi),
\label{eq:delta_sin}
\end{align}
where we used $R=\cos\phi$.

The instantaneous force-correlation coefficient is therefore
\begin{equation}
r(t)
=-\frac{1}{\sqrt{1+\Delta^2(t)/\sigma(t)^2}}.
\label{eq:r_instant}
\end{equation}

Introducing the dimensionless driving strength
\begin{equation}
Q\equiv\frac{A^2}{\sigma^2}=\frac{kA^2}{k_BT},
\end{equation}
we can write
\begin{equation}
r(t)=-\frac{1}{\sqrt{1+Q\sin^2\phi\,\sin^2(\omega t-\phi)}}.
\label{eq:r_with_Q}
\end{equation}
Here, $Q$ measures the driving strength relative to thermal fluctuations, and $\sigma^2 = k_BT/k$ denotes the steady-state variance of the harmonic trap, which remains time-independent under sinusoidal driving.
\\

\paragraph{Time-Averaged Force Correlation and Injected Power.}
For periodic driving, the force-correlation coefficient is naturally averaged over
one driving period $\tau=2\pi/\omega$,
\begin{equation}
\langle r\rangle_\tau=\frac{1}{\tau}\int_0^\tau r(t)\,dt.
\end{equation}
Introducing the phase variable $\theta=\omega t$, this becomes
\begin{equation}
\langle r\rangle_\tau
=-\frac{1}{2\pi}\int_0^{2\pi}
\frac{d\theta}{\sqrt{1+Q\,\sin^2\phi\,\sin^2(\theta-\phi)}} .
\end{equation}
Exploiting the periodicity and symmetry of the integrand, the integral reduces to
\begin{equation}
\langle r\rangle_\tau
=-\frac{2}{\pi}\int_0^{\pi/2}
\frac{d\theta}{\sqrt{1+Q\sin^2\phi\,\sin^2\theta}} .
\end{equation}

This time average isolates the geometric contribution of force alignment over a
driving cycle, independent of the instantaneous phase. Writing
\begin{equation}
\alpha \equiv Q\sin^2\phi ,
\label{eq:def_alpha}
\end{equation}
the integral can be expressed exactly in terms of the complete elliptic integral
of the first kind with negative parameter,
\begin{equation}
\langle r\rangle_\tau
= -\frac{2}{\pi}\,K(-\alpha),
\qquad
K(-\alpha)=\int_0^{\pi/2}\frac{d\theta}{\sqrt{1+\alpha\sin^2\theta}} ,
\label{eq:r_exact_sin}
\end{equation}
which is valid for all values of $Q$ and $\phi$.

Importantly, $\langle r\rangle_\tau$ by itself does not determine the entropy
production rate; it captures the geometric alignment contribution once the force
magnitudes are specified. In the harmonic-trap examples considered here, those
magnitudes are fixed or constrained by the same control parameters, so the
correlation becomes a direct proxy for entropy production only within that model
class.

For sinusoidal driving in this harmonic protocol, the time-averaged injected
mechanical power is governed by the same geometric parameter $\alpha$,
\begin{equation}
\langle P\rangle_\tau
= \frac{k\,k_BT}{2\gamma}\,\alpha .
\end{equation}
For completeness, we provide a short derivation of the expression below. The instantaneous injected power is
\begin{equation}
P(t)
= \left\langle \frac{\partial U(x,t)}{\partial t} \right\rangle
= -k\langle x(t)-\lambda(t)\rangle\,\dot{\lambda}(t)
= -k\,\Delta(t)\,\dot{\lambda}(t),
\end{equation}
with $\dot{\lambda}(t)=-A\omega\sin(\omega t)$ and
$\Delta(t)=A\sin\phi\,\sin(\omega t-\phi)$. Averaging over one period then gives
\begin{align}
\langle P\rangle_\tau &=
\frac{kA^2\omega}{2}\sin\phi\cos\phi \nonumber\\
&=
\frac{kA^2\omega}{2}\,
\frac{\omega/\Gamma}{1+(\omega/\Gamma)^2} 
=
\frac{k\,k_BT}{2\gamma}\,\alpha,
\end{align}
where we used $\Gamma=k/\gamma$, $\sigma^2=k_BT/k$, and
$\alpha=Q\sin^2\phi$ with $Q=A^2/\sigma^2$.
Thus, both force correlation and injected power are organized by the same
combination of driving amplitude and frequency, a consequence of the harmonic
structure of the protocol rather than a generic feature of driven systems.

\paragraph{Cycle-Averaged Entropy Production Rate.}
Because the net thermodynamic force is
\begin{align}
\vec F_{\rm net}(x,t)
&=
\vec F_{\rm ext}(x,t)+\vec F_{\rm info}(x,t) \nonumber\\
&=
k\bigl(\lambda(t)-\mu(t)\bigr)
=
-k\Delta(t),
\end{align}
the cycle-averaged entropy production rate follows from the second moment of the
lag:
\begin{align}
\langle \dot S_i\rangle_\tau
&=
\frac{D}{k_B T^2}
\left\langle |\vec F_{\rm net}|^2 \right\rangle_\tau =
\frac{Dk^2}{k_B T^2}
\left\langle \Delta^2(t) \right\rangle_\tau.
\end{align}
Using
\begin{equation}
\Delta(t)=A\sin\phi\,\sin(\omega t-\phi),
\end{equation}
and $\langle \sin^2(\omega t-\phi)\rangle_\tau = 1/2$, we obtain
\begin{equation}
\langle \Delta^2(t)\rangle_\tau
=
\frac{A^2\sin^2\phi}{2}.
\end{equation}
Hence,
\begin{equation}
\langle \dot S_i\rangle_\tau
=
\frac{k k_B}{2\gamma}\,\alpha.
\end{equation}

A natural reference point occurs at $\phi=\pi/4$, corresponding to driving at the
intrinsic relaxation rate $\omega=k/\gamma$.
Under this  condition, elastic and dissipative contributions to the response are equal, representing a crossover between quasi-static and dissipation-dominated operation.
The corresponding force correlation:
\begin{equation}
\langle r\rangle_\tau
= -\frac{2}{\pi}\,
K\!\left(-\frac{Q}{2}\right),
\end{equation}
depends smoothly on $Q$ and characterizes finite-power nonequilibrium operation.

\section{Mapping Experimental Data to the \\ Geometric Chart}

\label{AppenExpMapping}

This section details how experimental operating points are positioned within the
geometric design framework introduced in the main text.
We first consider the constant-velocity dragging experiment of
Di~Terlizzi \emph{et al.}~\cite{DiTerlizzi2024}, followed by the sinusoidal-driving
optical-tweezer microrheology experiment of
Catal{\`a}-Castro \emph{et al.}~\cite{CatalaCastro2025}.
In both cases, we explicitly state which quantities are preserved under rescaling
and how the force-correlation coefficient is affected. 
\\ 

\subsection*{Constant-velocity Dragging: Experimental Parameters }
The key parameters reported in Ref.~\cite{DiTerlizzi2024} are summarized in
Table~\ref{tab:exp_params}.
From these values, the steady-state input power and force correlation
$r_{\mathrm{ss}}$ follow directly.

\begin{table}[h!]
\centering
\caption{Parameters reported for the dragged harmonic trap experiment of
Di~Terlizzi \emph{et al.}, together with the inferred steady-state power and force correlation}
\label{tab:exp_params}
\begin{tabular}{lccc}
\toprule
\textbf{Parameter} & \textbf{Symbol} & \textbf{Value} & \textbf{Unit} \\
\midrule
Trap stiffness & $k_{\mathrm{exp}}$ & 70 & $\mathrm{pN}/\mu\mathrm{m}$ \\
Dragging speed & $v_{\mathrm{exp}}$ & 10 & $\mu\mathrm{m/s}$ \\
Mobility & $\mu_{\mathrm{exp}}$ & $4\times10^{4}$ & $\mathrm{nm/(pN\cdot s)}$ \\
Friction coefficient & $\gamma_{\mathrm{exp}}$ & 0.025 & $\mathrm{pN\cdot s}/\mu\mathrm{m}$ \\
\midrule
Input power & $P_{\mathrm{exp}}=\gamma_{\mathrm{exp}}v_{\mathrm{exp}}^2$ & 2.5 & $\mathrm{aW}$ \\
Correlation &
$r_{\mathrm{ss}}^{\mathrm{(exp)}}$ & $-0.91$ & -- \\
\multicolumn{4}{l}{\footnotesize (computed)} \\
\bottomrule
\end{tabular}
\end{table}

The experimental correlation follows from
\[
r_{\mathrm{ss}}
= -\left[1+\frac{(\gamma_{\mathrm{exp}}v_{\mathrm{exp}})^2}{k_{\mathrm{exp}}k_BT}\right]^{-1/2}.
\]

\begin{figure}[t!]
\centering
\includegraphics[width=0.5\textwidth]{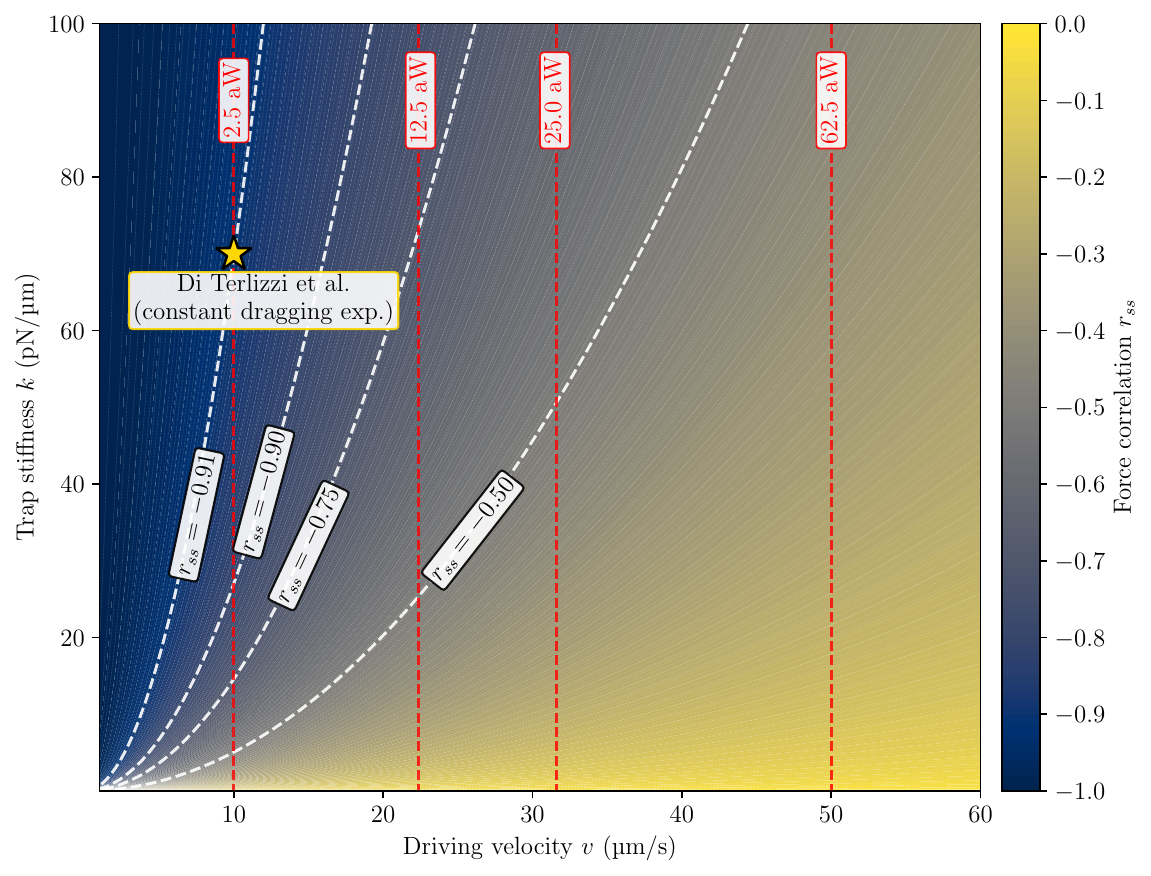} 
\caption{
\textbf{Geometric design chart matching experimental dissipation.}
Heat map of the steady-state force--correlation coefficient $r_{\mathrm{ss}}$ in the
$(k,v)$ parameter space of a harmonically trapped particle, computed at the experimental
friction coefficient $\gamma = 0.025\,\mathrm{pN\cdot s}/\mu\mathrm{m}$ reported by
Di~Terlizzi \emph{et al.}
Dashed white curves denote contours of constant force correlation, while dashed red curves
indicate contours of constant injected power $P=\gamma v^2$.
The yellow star marks the actual experimental operating point
$(k=70\,\mathrm{pN}/\mu\mathrm{m},\, v=10\,\mu\mathrm{m/s})$,
corresponding to $r_{\mathrm{ss}}\approx -0.91$ and
$P\approx 2.5\,\mathrm{aW}$.
A representative force--correlation contour
($r_{\mathrm{ss}}\approx -0.91$) is highlighted to guide comparison with the
rescaled geometric chart shown in the main text.
}
\label{fig:SI_design_chart_gamma025}
\end{figure}
To enable a direct comparison, the friction coefficient is fixed to its experimental value while the remaining control parameters are varied.
\\ 

\paragraph{Rescaling to a Reference Friction Coefficient.}
The geometric chart in the main text is constructed using a reference friction
coefficient $\gamma_0=0.01\,\mathrm{pN\cdot s}/\mu\mathrm{m}$, chosen for consistency
with the parameter values used in the main-text chart and related theoretical
studies. To place the experimental operating point on this chart, we rescale the driving
velocity while imposing:
\begin{enumerate}
\item Constant input power: $\gamma_0 v_0^2 = P_{\mathrm{exp}}$,
\item Constant trap stiffness: $k_0 = k_{\mathrm{exp}}$.
\end{enumerate}

Here we retain the original experimental friction coefficient
$\gamma_{\mathrm{exp}}=0.025\,\mathrm{pN\cdot s}/\mu\mathrm{m}$ so that the
experimental operating point is shown directly, while the main-text chart uses
the reference value $\gamma=0.01\,\mathrm{pN\cdot s}/\mu\mathrm{m}$ for visual
comparison.

The rescaled velocity is therefore
\begin{equation}
v_0
= v_{\mathrm{exp}}\sqrt{\frac{\gamma_{\mathrm{exp}}}{\gamma_0}}
\approx 15.8\,\mu\mathrm{m/s}.
\end{equation}
This rescaling preserves the experimentally relevant energy injection scale and
provides a convenient way to place the experimental point on the reference chart.
\\

\paragraph{Effect on Force Correlation.}
While this rescaling preserves the injected power, it does not preserve the
dimensionless group controlling the correlation,
\begin{equation}
r_{\mathrm{ss}}
= -\left(1+\frac{(\gamma v)^2}{k k_BT}\right)^{-1/2},
\end{equation}
where we use $k_BT\simeq4.1\,\mathrm{pN\cdot nm}$ corresponding to room temperature.

For the experimental parameters reported in Ref.~\cite{DiTerlizzi2024},
the force-correlation coefficient evaluates to
$r_{\mathrm{ss}}\approx -0.91$.
In Fig.~\ref{fig:SI_design_chart_gamma025}, this operating point is mapped onto a
reference chart constructed at $\gamma_0=0.01\,\mathrm{pN\cdot s/\mu m}$
by rescaling the velocity at fixed power and stiffness.
This rescaling shifts the corresponding correlation to
$r_{\mathrm{ss}}\approx -0.96$ while keeping the power scale fixed in the reference-chart construction.
The experimentally realized correlation remains $r_{\mathrm{ss}}\approx -0.91$.
\\

\paragraph{Geometric Chart at Experimental Friction.}
For direct comparison, Fig.~\ref{fig:SI_design_chart_gamma025} shows the geometric
design chart computed using the experimental friction coefficient
$\gamma_{\mathrm{exp}}=0.025\,\mathrm{pN\cdot s}/\mu\mathrm{m}$.
The experimental operating point lies close to the contour
$r_{\mathrm{ss}}=-0.95$, suggesting that the experimental point is compatible with a regime of strong force anti-alignment. This comparison indicates that the geometric design framework remains useful when evaluated directly at experimental parameter values, without any rescaling.
\\

\subsection{Sinusoidal Driving: Optical-Tweezer Microrheology}
We next consider the sinusoidal-driving experiment of
Catal{\`a}-Castro \emph{et al.}~\cite{CatalaCastro2025}.
Table~\ref{tab:exp_params_sin} lists the experimentally reported parameters
for the $f=100$\,Hz measurement performed in water.

\begin{table}[h!]
\centering
\caption{Parameters for the sinusoidal optical-tweezer experiment.}
\label{tab:exp_params_sin}
\begin{tabular}{lccc}
\toprule
\textbf{Parameter} & \textbf{Symbol} & \textbf{Value} & \textbf{Unit }\\
\midrule
Trap stiffness & $k$ & 50 & $\mathrm{pN}/\mu\mathrm{m}$ \\
Amplitude & $A$ & 200 & nm \\
Frequency & $f$ & 100 & Hz \\
Friction coefficient & $\gamma$ & 9.42 & $\mathrm{pN\cdot s}/\mu\mathrm{m}$ \\
\midrule
Dimensionless driving strength & $Q$ & 486.8 & -- \\
\multicolumn{4}{l}{\footnotesize (computed)} \\
\bottomrule
\end{tabular}
\end{table}

Within our framework, the experimental control parameters are naturally
organized through the dimensionless driving strength
\begin{equation}
Q = \frac{kA^2}{k_B T},
\end{equation}
which compares the elastic energy injected by the drive to the thermal energy scale.
Using the reported experimental values yields $Q = 486.8 \gg 1$,
placing the system deep in a strongly driven regime.

For sinusoidal driving in a harmonic trap, force geometry is controlled
by the driving strength together with the relative phase between the imposed drive
and the particle response. While this phase lag is not reported as an independent
experimental observable, the large ratio $\gamma\omega/k$ at the experimental
frequency implies a substantial lag characteristic of high-frequency
microrheological probing.

As a result, despite the large driving strength, the experiment is consistent with
a strongly driven regime with limited force alignment, rather than one evidently
optimized for energetic transduction. This behavior is reflected geometrically in
the chart by the experimental operating point at fixed $Q$.

\section{The Underdamped case: \\ A Supporting Argument}
\label{Appen_underdamped}

Here we provide a minimal analytical argument supporting the conjecture that
underdamped dynamics can sustain strong force anti-alignment while maintaining
finite currents, a situation that does not arise in overdamped systems.

We consider one-dimensional underdamped Langevin dynamics,
\begin{align}
\dot{x}_t &= \frac{p_t}{m}, \\
\dot{p}_t &= F_{\mathrm{ext}}(x_t,t) - \frac{\gamma}{m}p_t
+ \sqrt{2\gamma k_B T}\,\xi_t ,
\end{align}
where $F_{\mathrm{ext}}=-\partial_x U(x,t)$.
The phase-space density $\rho(x,p,t)$ obeys the Klein--Kramers equation.

In general, $\rho(x,p,t)$ does not factorize and both information forces
depend on both position and momentum.
Exact factorization of the phase-space distribution,
$\rho(x,p,t)=\rho_x(x,t)\rho_p(p,t)$, occurs only under special conditions
(e.g.\ equilibrium) or when enforced by additional constraints discussed below.

We define position- and momentum-space information geometric quantities:
\begin{align}
F_{\mathrm{info}}^{(x)} &= -k_B T\,\partial_x \ln \rho, \\
F_{\mathrm{info}}^{(p)} &= -k_B T\,\partial_p \ln \rho,
\end{align}
which are defined on full phase space. The dimension of $F_{\mathrm{info}}^{(x)}$ is that of a force, $MLT^{-2}$,
whereas $F_{\mathrm{info}}^{(p)}$ has the dimension of a velocity, $LT^{-1}$. Upon conditional averaging over momentum, the position-space information force reduces to the information force associated with the position marginal:
\begin{align}
\overline{F}_{\mathrm{info}}^{(x)}(x,t)
&:=
\int dp\,\rho(p|x,t)\,
F_{\mathrm{info}}^{(x)}(x,p,t) \nonumber\\
&=
-k_B T\,\partial_x\ln \rho_x(x,t).
\end{align}
\subsection*{Perfect Position-Space Anti-Alignment}

Assume that the external force is perfectly anti-aligned with the
position-space information force in the sense of proportional anti-alignment,
\begin{equation}
F_{\mathrm{ext}}(x,t)
=
-c\,F_{\mathrm{info}}^{(x)}(x,p,t),
\qquad c>0,
\label{eq:underdamped_align}
\end{equation}
for all $(x,p)$ with $\rho>0$.
The special case $c=1$ corresponds to exact pointwise force cancellation.
This condition is highly restrictive and is not expected to hold
generically in driven underdamped systems.

Since $F_{\mathrm{ext}}$ is independent of $p$, this condition implies that
$\partial_x \ln \rho$ is independent of $p$, and therefore
\begin{equation}
\rho(x,p,t)=\rho_x(x,t)\rho_p(p,t).
\end{equation}
Equation~\eqref{eq:underdamped_align} then yields
\begin{equation}
\rho_x(x,t)
\propto
\exp\left[-\frac{U(x,t)}{ck_B T}\right].
\end{equation}
Thus, perfect proportional anti-alignment in position space enforces a
Boltzmann-like structure in the positional sector, with an effective
temperature $T_{\rm eff}=cT$. The special case $c=1$ corresponds to
exact pointwise force cancellation and recovers the equilibrium positional
distribution at temperature $T$. We emphasize that this is a kinematic consequence of imposing Eq.~\eqref{eq:underdamped_align} pointwise at each instant; it characterizes the distributional shape enforced by exact force anti-alignment but does not by itself assert that such $\rho_x(x,t)$ is dynamically self-sustained under the full Klein-Kramers evolution.

Thus, perfect proportional force anti-alignment in position space constrains
only the positional sector, without directly constraining the momentum
distribution.

\subsection*{Entropy Production and Current}

For standard underdamped Langevin dynamics with linear friction and a single
thermal reservoir, the entropy production rate is entirely determined by the
irreversible momentum current~\cite{Seifert2012Stochastic,SpinneyFord2012}:
\begin{equation}
\dot{S}_i(t)=\frac{\gamma}{ T}\iint \rho(x,p,t)
\Bigl(\frac{p}{m}-F_{\mathrm{info}}^{(p)}\Bigr)^2\,dx\,dp .
\end{equation}
If the phase-space distribution factorizes,
$\rho(x,p,t)=\rho_x(x,t)\rho_p(p,t)$, as implied by the strong condition of
perfect position-space force anti-alignment, this reduces to
\begin{equation}
\dot{S}_i(t)
=
\frac{\gamma}{T}
\Bigl\langle
\Bigl(
\frac{p}{m}-F_{\mathrm{info}}^{(p)}
\Bigr)^2
\Bigr\rangle_p .
\label{eq:EPR_underdamped_reduced}
\end{equation}
Hence, imposing perfect force anti-alignment in position space places no
direct constraint on entropy production, which remains generically nonzero
unless the momentum distribution is the equilibrium Maxwellian at temperature
$T$ with zero mean momentum.

Meanwhile, in the overdamped case, the mean transport velocity is
$d_t\langle x\rangle=\int dx\,J(x,t)=\mu\langle F_{\rm net}\rangle$,
where $J(x,t)=\mu\rho(x,t)F_{\rm net}(x,t)$ is the local probability current.
In underdamped dynamics, by contrast, the corresponding mean transport velocity is
$d_t\langle x\rangle=\langle p\rangle/m$, and is therefore fixed by the first moment
of the momentum marginal. Thus, even when forces are perfectly anti-aligned in
position space, finite mean transport can persist if the momentum sector is
driven or otherwise constrained away from equilibrium.

Crucially, the mean current is governed by the first moment of the momentum
distribution, whereas the entropy production depends on the quadratic
combination
$\left\langle\left(p/m-F_{\mathrm{info}}^{(p)}\right)^2\right\rangle_p$.

Equation~\eqref{eq:EPR_underdamped_reduced}
shows that, at the level of the phase-space force decomposition, perfect
force anti-alignment in position space does not preclude finite directed
current, provided the momentum sector is driven
or otherwise constrained away from equilibrium. This particular coexistence does not arise in overdamped systems, where
global force anti-alignment, $r=-1$, implies a vanishing mean current under
standard boundary conditions. In overdamped dynamics, entropy production
vanishes only under exact pointwise force cancellation,
$\vec F_{\rm ext}+\vec F_{\rm info}=0$, while finite entropy production may
persist at stall in the absence of transport, thereby precluding finite power
operation. 

Therefore, in underdamped dynamics, momentum acts as a buffer that decouples
position-space force geometry from transport and from the direct
entropy-production channel. The relevant  geometry here is instead the
velocity-space geometry in the momentum sector. This observation supports the
conjecture that inertia could be exploited as a design resource for sustaining
finite currents while maintaining near-cancellation in the positional force
sector of underdamped stochastic machines.

\bibliographystyle{unsrtnat}
\bibliography{references}

\begin{thebibliography}{26}
\providecommand{\natexlab}[1]{#1}
\providecommand{\url}[1]{\texttt{#1}}
\expandafter\ifx\csname urlstyle\endcsname\relax
  \providecommand{\doi}[1]{doi: #1}\else
  \providecommand{\doi}{doi: \begingroup \urlstyle{rm}\Url}\fi

\bibitem[Sekimoto(1998)]{Sekimoto1998Langevin}
K.~Sekimoto.
\newblock Langevin {E}quation and {T}hermodynamics.
\newblock \emph{Prog. Theor. Phys. Suppl.}, 130:\penalty0 17--27, 1998.
\newblock \doi{10.1143/PTPS.130.17}.

\bibitem[Di~Terlizzi et~al.(2024)Di~Terlizzi, Gironella, Herraez-Aguilar, Betz,
  Monroy, Baiesi, and Ritort]{DiTerlizzi2024}
I.~Di~Terlizzi, M.~Gironella, D.~Herraez-Aguilar, T.~Betz, F.~Monroy,
  M.~Baiesi, and F.~Ritort.
\newblock Variance sum rule for entropy production.
\newblock \emph{Science}, 383\penalty0 (6686):\penalty0 971--976, 2024.
\newblock \doi{10.1126/science.adh1823}.

\bibitem[Di~Terlizzi(2025)]{DiTerlizzi2025}
I.~Di~Terlizzi.
\newblock Force-{F}ree {K}inetic {I}nference of {E}ntropy {P}roduction.
\newblock \emph{Phys. Rev. Lett.}, 135:\penalty0 237101, 2025.
\newblock \doi{10.1103/fsph-437v}.

\bibitem[Barato and Seifert(2015)]{Barato2015Thermodynamic}
A.~C. Barato and U.~Seifert.
\newblock Thermodynamic {U}ncertainty {R}elation for {B}iomolecular
  {P}rocesses.
\newblock \emph{Phys. Rev. Lett.}, 114:\penalty0 158101, 2015.
\newblock \doi{10.1103/PhysRevLett.114.158101}.

\bibitem[Gingrich et~al.(2016)Gingrich, Horowitz, Perunov, and
  England]{Gingrich2016Dissipation}
T.~R. Gingrich, J.~M. Horowitz, N.~Perunov, and J.~L. England.
\newblock Dissipation {B}ounds {A}ll {S}teady-{S}tate {C}urrent {F}luctuations.
\newblock \emph{Phys. Rev. Lett.}, 116:\penalty0 120601, 2016.
\newblock \doi{10.1103/PhysRevLett.116.120601}.

\bibitem[Seifert(2019)]{Seifert2019TUR}
U.~Seifert.
\newblock From {S}tochastic {T}hermodynamics to {T}hermodynamic {I}nference.
\newblock \emph{Annu. Rev. Condens. Matter Phys.}, 10:\penalty0 171--192, 2019.
\newblock \doi{10.1146/annurev-conmatphys-031218-013554}.

\bibitem[Horowitz and Gingrich(2020)]{horowitz2020thermodynamic}
J.~M. Horowitz and T.~R. Gingrich.
\newblock Thermodynamic uncertainty relations constrain non-equilibrium
  fluctuations.
\newblock \emph{Nat. Phys.}, 16\penalty0 (1):\penalty0 15--20, 2020.
\newblock \doi{10.1038/s41567-019-0702-6}.

\bibitem[Nicholson et~al.(2020)Nicholson, Garc{\'\i}a-Pintos, del Campo, and
  Green]{Nicholson2020TimeInformation}
S.~B. Nicholson, L.~P. Garc{\'\i}a-Pintos, A.~del Campo, and J.~R. Green.
\newblock Time--information uncertainty relations in thermodynamics.
\newblock \emph{Nat. Phys.}, 16:\penalty0 1211--1215, 2020.
\newblock \doi{10.1038/s41567-020-0981-y}.

\bibitem[Dechant and Sasa(2021)]{DechantSasa2021PhysRevX11.041061}
A.~Dechant and S.-i. Sasa.
\newblock Improving {T}hermodynamic {B}ounds {U}sing {C}orrelations.
\newblock \emph{Phys. Rev. X}, 11:\penalty0 041061, 2021.
\newblock \doi{10.1103/PhysRevX.11.041061}.

\bibitem[Yang and Qian(2021)]{yang2021bivectorial}
Ying-Jen Yang and Hong Qian.
\newblock Bivectorial nonequilibrium thermodynamics: Cycle affinity, vorticity
  potential, and onsager’s principle.
\newblock \emph{J. Stat. Phys.}, 182\penalty0 (3):\penalty0 46, 2021.

\bibitem[O'Byrne(2023)]{Byrne2023}
J.~O'Byrne.
\newblock Nonequilibrium currents in stochastic field theories: A geometric
  insight.
\newblock \emph{Phys. Rev. E}, 107:\penalty0 054105, May 2023.
\newblock \doi{10.1103/PhysRevE.107.054105}.

\bibitem[Risken(1996)]{Risken1989Fokker}
H~Risken.
\newblock \emph{The Fokker-Planck Equation: Methods of Solution and
  Applications}.
\newblock Springer Series in Synergetics. Springer, Berlin, Heidelberg, 2
  edition, 1996.
\newblock ISBN 978-3-642-61544-3.
\newblock \doi{10.1007/978-3-642-61544-3}.

\bibitem[Seifert(2005)]{Seifert2005Entropy}
U.~Seifert.
\newblock Entropy {P}roduction along a {S}tochastic {T}rajectory and an
  {I}ntegral {F}luctuation {T}heorem.
\newblock \emph{Phys. Rev. Lett.}, 95:\penalty0 040602, 2005.
\newblock \doi{10.1103/PhysRevLett.95.040602}.

\bibitem[Glansdorff and Prigogine(1971)]{glansdorff1971thermodynamic}
P.~Glansdorff and I.~Prigogine.
\newblock \emph{Thermodynamic Theory of Structure, Stability and Fluctuations}.
\newblock Wiley-Interscience, New York, 1971.
\newblock ISBN 978-0471302803.

\bibitem[Jarzynski(1997)]{Jarzynski1997Nonequilibrium}
C.~Jarzynski.
\newblock Nonequilibrium {E}quality for {F}ree {E}nergy {D}ifferences.
\newblock \emph{Phys. Rev. Lett.}, 78:\penalty0 2690--2693, 1997.
\newblock \doi{10.1103/PhysRevLett.78.2690}.

\bibitem[Seifert(2012)]{Seifert2012Stochastic}
U.~Seifert.
\newblock Stochastic thermodynamics, fluctuation theorems and molecular
  machines.
\newblock \emph{Rep. Prog. Phys.}, 75:\penalty0 126001, 2012.
\newblock \doi{10.1088/0034-4885/75/12/126001}.

\bibitem[Neuman and Block(2004)]{Neuman2004OpticalTrapping}
K.~C. Neuman and S.~M. Block.
\newblock Optical trapping.
\newblock \emph{Rev. Sci. Instrum.}, 75:\penalty0 2787--2809, 2004.
\newblock \doi{10.1063/1.1785844}.

\bibitem[Kumar et~al.(2021)Kumar, Vitali, Wiedemann, Quagliarini, Ferracci,
  Cordella, Ponzini, Zaccarelli, Roca-Bonet, Pesce, De~Luca, Marag{\`o}, and
  Volpe]{Kumar2021Multi}
R.~Kumar, V.~Vitali, T.~Wiedemann, A.~Quagliarini, R.~Ferracci, F.~Cordella,
  M.~Ponzini, E.~Zaccarelli, J.~Roca-Bonet, G.~Pesce, A.~C. De~Luca, O.~M.
  Marag{\`o}, and G.~Volpe.
\newblock Multi-frequency passive and active microrheology with optical
  tweezers.
\newblock \emph{Sci. Rep.}, 11:\penalty0 13917, 2021.
\newblock \doi{https://doi.org/10.1038/s41598-021-93130-x}.

\bibitem[Catal{\`a}-Castro et~al.(2025)Catal{\`a}-Castro, Ortiz-V{\'a}squez,
  Mart{\'i}nez-Fern{\'a}ndez, Pezzano, Garcia-Cabau, Fern{\'a}ndez-Campo,
  Sanfeliu-Cerd{\'a}n, Jim{\'e}nez-Delgado, Salvatella, Ruprecht, Frigeri, and
  Krieg]{CatalaCastro2025}
F.~Catal{\`a}-Castro, S.~Ortiz-V{\'a}squez, C.~Mart{\'i}nez-Fern{\'a}ndez,
  F.~Pezzano, C.~Garcia-Cabau, M.~Fern{\'a}ndez-Campo, N.~Sanfeliu-Cerd{\'a}n,
  S.~Jim{\'e}nez-Delgado, X.~Salvatella, V.~Ruprecht, P.-A. Frigeri, and
  M.~Krieg.
\newblock Measuring age-dependent viscoelasticity of organelles, cells and
  organisms with time-shared optical tweezer microrheology.
\newblock \emph{Nat. Nanotechnol.}, 20:\penalty0 411--420, 2025.
\newblock \doi{10.1038/s41565-024-01830-y}.

\bibitem[Note1()]{Note1}
Note1.
\newblock This mapping should be interpreted as a placement of an
  experimentally realized constant-drag protocol within the force-geometric
  chart, not as a direct model of the entire results of the RBC
  membrane-flickering experiment in Ref.~\cite {DiTerlizzi2024}.

\bibitem[Ito and Dechant(2020)]{Ito2020Geometric}
S.~Ito and A.~Dechant.
\newblock Stochastic {T}ime {E}volution, {I}nformation {G}eometry, and the
  {C}ram\'er-{R}ao {B}ound.
\newblock \emph{Phys. Rev. X}, 10:\penalty0 021056, 2020.
\newblock \doi{10.1103/PhysRevX.10.021056}.

\bibitem[Dago et~al.(2021)Dago, Pereda, Barros, Ciliberto, and
  Bellon]{Ludovic2021}
S.~Dago, J.~Pereda, N.~Barros, S.~Ciliberto, and L.~Bellon.
\newblock Information and {T}hermodynamics: Fast and {P}recise {A}pproach to
  {Landauer}'s {B}ound in an {U}nderdamped {M}icromechanical {O}scillator.
\newblock \emph{Phys. Rev. Lett.}, 126:\penalty0 170601, 2021.
\newblock \doi{10.1103/PhysRevLett.126.170601}.

\bibitem[Oikawa et~al.(2025)Oikawa, Nakayama, Ito, and Bechhoefer]{Oikawa2025}
S.~Oikawa, Y.~Nakayama, S.~Ito, and J.~Bechhoefer.
\newblock Experimentally achieving minimal dissipation via thermodynamically
  optimal transport.
\newblock \emph{Nat. Commun.}, 16:\penalty0 10424, 2025.
\newblock \doi{10.1038/s41467-025-66519-9}.

\bibitem[van Kampen(2007)]{vanKampen2007Stochastic}
N.~G. van Kampen.
\newblock \emph{Stochastic Processes in Physics and Chemistry}.
\newblock Elsevier, 3 edition, 2007.

\bibitem[Sekimoto(2010)]{Sekimoto2010Stochastic}
K.~Sekimoto.
\newblock \emph{Stochastic Energetics}, volume 799 of \emph{Lecture Notes in
  Physics}.
\newblock Springer, 2010.
\newblock ISBN 978-3-642-05410-5.
\newblock \doi{10.1007/978-3-642-05411-2}.

\bibitem[Spinney and Ford(2012)]{SpinneyFord2012}
R.~E. Spinney and I.~J. Ford.
\newblock Entropy production in full phase space for continuous stochastic
  dynamics.
\newblock \emph{Phys. Rev. E}, 85:\penalty0 051113, 2012.
\newblock \doi{10.1103/PhysRevE.85.051113}.

\end{thebibliography}

\end{document}